\documentclass{article}%
\usepackage{amssymb}
\usepackage{amsmath}
\usepackage{hyperref}
\usepackage{amsfonts}
\usepackage{graphicx}%
\providecommand{\U}[1]{\protect\rule{.1in}{.1in}}
\begin{document}

\title{Grounded Hyperspheres as Squashed Wormholes}
\author{H Alshal$^{1,2}$\thanks{{\small alshal@cu.edu.eg}} \ and T Curtright$^{2}%
$\thanks{{\small curtright@miami.edu}}\\$^{1}$Department of Physics, Cairo University, Giza, 12613, Egypt\\$^{2}$Department of Physics, University of Miami, Coral Gables, FL 33124-8046, USA}
\date{5 December 2018}
\maketitle

\begin{abstract}
We compute exterior Green functions for equipotential, grounded\ hyperspheres
in $N$-dimensional electrostatics by squashing Riemannian wormholes,\ where an
image charge is placed in the branch of the wormhole opposite the branch
containing the source charge, thereby providing a vivid geometrical approach
to a method first suggested in 1897 by Sommerfeld. \ We compare and contrast
the strength and location of the image charge in the wormhole approach with
that of the conventional Euclidean solution where an image charge of reduced
magnitude is located inside the hypersphere. \ While the two approaches give
mathematically equivalent Green functions, we believe they provide strikingly
different physics perspectives.

\end{abstract}

\begin{center}
\bigskip

In tribute to Richard Feynman (1918-1988) and Arnold Sommerfeld (1868-1951)
\end{center}

\section{Introduction}

\href{https://en.wikipedia.org/wiki/Richard_Feynman}{Feynman} constantly
emphasized that insight could be gained by approaching a problem from a
different point of view \cite{Mehra}. \ We follow that philosophy here to
construct a Green function, $G_{o}$,\ for the $N$-dimensional electrostatics
of an equipotential, grounded hypersphere --- i.e. a \textquotedblleft
conducting\textquotedblright\ hypersphere. \ We stress the geometrical aspects
of a method first employed in the late 19th century by
\href{https://en.wikipedia.org/wiki/Arnold_Sommerfeld}{Sommerfeld}
\cite{Eckert}, albeit not for this specific problem. \ Although the method was
introduced some 120 years ago \cite{Sommerfeld}, we believe it suggests
insights that are not widely appreciated. \ While Sommerfeld's method has been
employed during the intervening century to solve a handful of otherwise
difficult electrostatic and heat conduction problems
\cite{Hobson,Waldmann,DavisReitz,DavisReitzAgain,Duffy}, we believe that a
stronger emphasis on its geometrical aspects is worthwhile and justifies
applying the method to a broader class of problems, even those which are not
difficult to solve by other means. \ To that end we reconsider grounded
hyperspheres in $N$ \emph{spatial} dimensions.

Although the Green function with homogeneous Dirichlet boundary conditions is
known for the grounded hypersphere and can be obtained using well-known
techniques, here we compute $G_{o}$ by a novel method that uses
\textquotedblleft wormholes\textquotedblright\ \cite{MorrisThorne,JamesEtAl}%
\ in an $N$-dimensional Riemannian space. \ We build appropriate Green
functions for the invariant Laplacian, $\nabla^{2}$, acting on various
wormhole geometries, initially by imposing boundary conditions only
asymptotically, to obtain $G$, and finally by requiring that the Green
function also vanish at the narrowest part of the wormhole's \textquotedblleft
throat\textquotedblright\ to obtain $G_{o}$. \ 

To be more specific, we consider $N$-dimensional versions of the isotropic
Ellis wormhole \cite{Ellis}, whose equatorial slices have radii given by
$r\left(  w\right)  =\sqrt{R^{2}+w^{2}}$ where $R$ is a constant and
$-\infty\leq w\leq+\infty$. \ We construct these manifolds so that the region
near $w=0$ is a curved bridge\ \cite{EinsteinRosen}\ that connects two
distinct branches of the manifold, with those branches approaching two
separate copies of $N$-dimensional Euclidean space, $\mathbb{E}_{N}$,
asymptotically as $w\rightarrow\pm\infty$ (please see the Figures). \ We build
appropriate Green functions for $\nabla^{2}$ acting on these specific
geometries, once again by first imposing boundary conditions only as
$w\rightarrow\pm\infty$, to obtain $G$, and then by requiring that the Green
function also vanish at $w=0$, i.e. at radius $R$, to obtain $G_{o}$.

The \textquotedblleft grounded\textquotedblright\ Green function $G_{o}$ is
constructed by placing a source and its negative image at exactly the
\emph{same radius} and in precisely the \emph{same direction} on the manifold,
but on \emph{opposite} branches --- somewhat striking but nevertheless very
natural positions in this context. \ 

Next we introduce deformations of the Ellis wormhole whose radii are given by
the $p$-norm $r\left(  w\right)  =\left(  R^{p}+\left\vert w\right\vert
^{p}\right)  ^{1/p}$. \ We compute Green functions $G$ and $G_{o}$ for this
family of manifolds as well. \ Finally, we consider the $p\rightarrow1$ limit
of this family of manifolds and Green functions. \ 

In the $p\rightarrow1$ limit, where $r\left(  w\right)  =R+\left\vert
w\right\vert $ is the so-called Manhattan\ norm, both branches of the manifold
are completely \textquotedblleft squashed flat\textquotedblright. \ That is to
say, the manifold degenerates into two distinct Euclidean spaces joined
together along a hyperspherical \textquotedblleft doorway\textquotedblright%
\ of radius $R$. (For equatorial slices of the manifolds as $p$ approaches
$1$, please see the Figures.) \ However, the \emph{interior} of the
hypersphere is \emph{excluded} from either branch. \ Moreover, in this limit
it is clear that by restricting $G_{o}$ to have source and field points on
just one of the distinct branches, a Green function for the grounded
equipotential hypersphere is obtained. \ Minor rearrangements of the terms
shows that $G_{o}$ is in exact agreement with the usual symmetrized Green
function for the grounded hypersphere, as obtained by considering a single
copy of $\mathbb{E}_{N}$ with an image charge placed inside the hypersphere.

\section{Electrostatics in $N$ Euclidean dimensions}

The point-particle electric potential in an $N$-dimensional Euclidean space,
$\mathbb{E}_{N}$, is well-known to vary inversely with distance as the
$\left(  N-2\right)  $-th power. \ For a unit point charge located at the
origin,
\begin{equation}
\Phi_{\mathbb{E}_{N}}\left(  \overrightarrow{r}\right)  =\frac{k_{N}}{r^{N-2}%
}\ ,\ \ \ \ \ \nabla^{2}\Phi_{\mathbb{E}_{N}}\left(  \overrightarrow{r}%
\right)  =-~\delta^{N}\left(  \overrightarrow{r}\right)  \ ,\ \ \ \ k_{N}%
\equiv\frac{1}{\left(  N-2\right)  \Omega_{N}}\label{PointCharge}%
\end{equation}
The total hyper-angle $\Omega_{N}$ (i.e. the area of the \emph{unit radius}
sphere, $S_{N-1}$, embedded in $N$ dimensions) is given by
\begin{equation}
\Omega_{N}=\int_{S_{N-1}}d\Omega=\frac{2\pi^{N/2}}{\Gamma\left(  N/2\right)
}\ .\label{HyperAngle}%
\end{equation}
where $d\Omega$ is the standard measure on $S_{N-1}$. \ For example,
$\Omega_{1}=2$, $\Omega_{2}=2\pi$, $\Omega_{3}=4\pi$, $\Omega_{4}=2\pi^{2}$,
etc. \ The case $N=2$ is handled as a limit, to obtain%
\begin{equation}
\Phi_{\mathbb{E}_{2}}\left(  \overrightarrow{r}\right)  =-\frac{1}{2\pi}%
\ln\left(  r/R\right)  \ ,\text{\ \ \ \ \ }\nabla^{2}\Phi_{\mathbb{E}_{2}%
}\left(  \overrightarrow{r}\right)  =-~\delta^{2}\left(  \overrightarrow{r}%
\right)  \ ,\label{PointOnSphere}%
\end{equation}
up to a constant $R$ that sets the distance scale. \ For more details in this
particular case, see \cite{CurtrightEtAl}.

Consequently, a Green function for $\mathbb{E}_{N}$ is
\begin{equation}
G_{\mathbb{E}_{N}}\left(  \overrightarrow{r_{1}};\overrightarrow{r_{2}%
}\right)  =\frac{k_{N}}{\left\vert \overrightarrow{r_{1}}%
-\overrightarrow{r_{2}}\right\vert ^{N-2}}\ ,\ \ \ \ \ \nabla^{2}%
G_{\mathbb{E}_{N}}\left(  \overrightarrow{r_{1}};\overrightarrow{r_{2}%
}\right)  =-~\delta^{N}\left(  \overrightarrow{r_{1}}-\overrightarrow{r_{2}%
}\right)  \label{EuclideanGreen}%
\end{equation}
This result is translationally invariant, that is to say, $G_{\mathbb{E}_{2}%
}\left(  \overrightarrow{r_{1}};\overrightarrow{r_{2}}\right)  $ depends only
on the difference $\overrightarrow{r_{1}}-\overrightarrow{r_{2}}$. \ Moreover,
this choice for the Green function incorporates boundary conditions at spatial
infinity that mimic the behavior in (\ref{PointCharge}). \ So any sufficiently
localized charge distribution $\rho\left(  \overrightarrow{r}\right)  $ gives
rise to the usual linear superposition,
\begin{equation}
\Phi\left(  \overrightarrow{r_{1}}\right)  =\int G_{\mathbb{E}_{N}}\left(
\overrightarrow{r_{1}};\overrightarrow{r_{2}}\right)  ~\rho\left(
\overrightarrow{r_{2}}\right)  ~d^{N}r_{2}\label{LinearSuperposition}%
\end{equation}
where we assume the integral is well-defined and finite. \ For a localized
charge distribution, $\Phi\left(  \overrightarrow{r_{1}}\right)
\underset{r_{1}\rightarrow\infty}{\sim}\frac{k_{N}Q}{r_{1}^{N-2}}+O\left(
\frac{1}{r_{1}^{N-1}}\right)  $ where $Q=\int\rho\left(  \overrightarrow{r_{2}%
}\right)  ~d^{2}r_{2}$ is the total charge. \ Finally, we note that the Green
function (\ref{EuclideanGreen}) in $N$ dimensions can be expanded in terms of
\href{https://en.wikipedia.org/wiki/Gegenbauer_polynomials}{Gegenbauer
polynomials}, $C_{l}^{\left(  \alpha\right)  }\left(  \cos\theta\right)  $,
namely,%
\begin{equation}
\frac{1}{\left\vert \overrightarrow{r_{1}}-\overrightarrow{r_{2}}\right\vert
^{N-2}}=\sum_{l=0}^{\infty}\frac{\left(  r_{<}\right)  ^{l}}{\left(
r_{>}\right)  ^{l+N-2}}~C_{l}^{\left(  \frac{N-2}{2}\right)  }\left(
\widehat{r_{1}}\cdot\widehat{r_{2}}\right)  \label{GegenbauerExpansion}%
\end{equation}
where $\widehat{r_{1}}\cdot\widehat{r_{2}}\equiv\cos\theta$ and $r_{>}$ or
$r_{<}$ is the $\max$ or $\min$ of $r_{1}$ and $r_{2}$, respectively. \ This
is a straightforward generalization of the well-known $N=3$ case that involves
the \href{https://en.wikipedia.org/wiki/Legendre_polynomials}{Legendre
polynomials} $P_{l}\left(  \cos\theta\right)  $.

\section{Electrostatics in $N$ curved dimensions}

On a Riemannian manifold, described by a metric $g_{\mu\nu}$, distance
increments are given by $ds$ where%
\begin{equation}
\left(  ds\right)  ^{2}=g_{\mu\nu}~dx^{\mu}dx^{\nu}%
\end{equation}
Summation over integer $\mu$ and $\nu$ is implicitly understood, with
$1\leq\mu,\nu\leq N$ for an $N$-dimensional manifold. \ The invariant
Laplacian on such a manifold is given by
\begin{equation}
\nabla^{2}=\frac{1}{\sqrt{g}}~\partial_{\mu}\left(  \sqrt{g}~g^{\mu\nu
}\partial_{\nu}\right)
\end{equation}
where $g\equiv\det g_{\mu\nu}$ and $g^{\mu\nu}$ is the matrix inverse of
$g_{\mu\nu}$. \ 

Consider now an infinite isotropic manifold with%
\begin{equation}
\left(  ds\right)  ^{2}=\left(  dw\right)  ^{2}+r^{2}\left(  w\right)  \left(
d\widehat{r}\right)  ^{2} \label{GeneralIsotropicWormhole}%
\end{equation}
The variable $w$ takes on values $-\infty\leq w\leq+\infty$, the unit vectors
$\widehat{r}$ represent the points on $S_{N-1}$ in terms of the standard
angular parameterization, and $r\left(  w\right)  $ is assumed to be a
positive, non-vanishing \textquotedblleft radius\textquotedblright\ function
that has a minimum at $w=0$ and becomes infinite as $w\rightarrow\pm\infty$.
\ For fixed angles, radial displacements on the manifold are determined just
by $ds=\pm dw$. \ 

We refer to $w>0$ and $w<0$ as the \textquotedblleft upper\textquotedblright%
\ and \textquotedblleft lower\textquotedblright\ branches of the manifold,
respectively, and following Einstein and Rosen \cite{EinsteinRosen}, we call
the region near $w=0$ the \textquotedblleft bridge\textquotedblright\ between
the two branches. \ For visualization purposes, please see the examples shown
in the Figures.

The metric, its inverse, and the determinant $g$ have well-known, standard
expressions for this manifold. \ In any case, the form of the metric leads to
the invariant Laplacian
\begin{equation}
\nabla^{2}=\frac{1}{r\left(  w\right)  ^{N-1}}~\partial_{w}\left(  r\left(
w\right)  ^{N-1}\partial_{w}\right)  -\frac{1}{r\left(  w\right)  ^{2}}%
~L^{2}\ ,\ \ \
\end{equation}
where all the $N-1$ angular derivatives are contained in $L_{jk}%
\equiv-i\left(  x_{j}\partial_{k}-x_{k}\partial_{j}\right)  $ with
$L^{2}\equiv\sum_{1\leq j<k\leq N}L_{jk}L_{jk}$. \ 

The eigenfunctions of $L^{2}$ form a complete set on $S_{N-1}$, the
\href{https://en.wikipedia.org/wiki/Spherical_harmonics#Higher_dimensions}{hyperspherical
harmonics} $Y_{lm_{1}m_{2}\cdots m_{N-2}}$, analogous to the familiar
\href{https://en.wikipedia.org/wiki/Spherical_harmonics}{spherical harmonics}
$Y_{lm}$ on $S_{2}$. \ The $Y_{lm_{1}m_{2}\cdots m_{N-2}}$ depend on the $N-1$
angles parameterizing all points on the hypersphere, but do not depend on $w$.
\ Thus, with unit vectors $\widehat{r_{1}}$ and $\widehat{r_{2}}$ representing
two points on $S_{N-1}$,
\begin{equation}
\sum_{l,m_{1},m_{2},\cdots,m_{N-2}}Y_{lm_{1}m_{2}\cdots m_{N-2}}\left(
\widehat{r_{1}}\right)  Y_{lm_{1}m_{2}\cdots m_{N-2}}^{\ast}\left(
\widehat{r_{2}}\right)  =\delta^{N-1}\left(  \widehat{r_{1}}-\widehat{r_{2}%
}\right)  \ ,\ \ \ \int\delta^{N-1}\left(  \widehat{r_{1}}-\widehat{r_{2}%
}\right)  d\Omega\left(  \widehat{r_{1}}\right)  =1
\end{equation}
More details about hyperspherical harmonics may be found in
\href{https://www.amazon.com/s/ref=nb_sb_noss_2?url=search-alias%3Dstripbooks&field-keywords=Hyperspherical+Harmonics}{various
books}, although notation and conventions may differ from those used here.
\ Acting on $Y_{lm_{1}m_{2}\cdots m_{N-2}}$ the $L^{2}\ $eigenvalues are given
by
\begin{equation}
L^{2}Y_{lm_{1}m_{2}\cdots m_{N-2}}=l\left(  l+N-2\right)  Y_{lm_{1}m_{2}\cdots
m_{N-2}}%
\end{equation}
for $l=0,1,2,\cdots$, generalizing the well-known $N=3$ case. \ As a further
generalization of $N=3$ results, there is an addition formula for
hyperspherical harmonics resulting in a Gegenbauer polynomial. \ For fixed $l$
in $N$ dimensions,%
\begin{equation}
\frac{2l+N-2}{N-2}~C_{l}^{\left(  \frac{N-2}{2}\right)  }\left(
\widehat{r_{1}}\cdot\widehat{r_{2}}\right)  =\Omega_{N}\sum_{m_{1}%
,m_{2},\cdots,m_{N-2}}Y_{lm_{1}m_{2}\cdots m_{N-2}}\left(  \widehat{r_{1}%
}\right)  Y_{lm_{1}m_{2}\cdots m_{N-2}}^{\ast}\left(  \widehat{r_{2}}\right)
\end{equation}
where our orthonormalization convention is
\begin{equation}
\int Y_{lm_{1}\cdots m_{N-2}}\left(  \widehat{r}\right)  Y_{l^{\prime}%
m_{1}^{\prime}\cdots m_{N-2}^{\prime}}^{\ast}\left(  \widehat{r}\right)
~d\Omega=\delta_{ll^{\prime}}\delta_{m_{1}m_{1}^{\prime}}~\cdots
~\delta_{m_{N-2}m_{N-2}^{\prime}}%
\end{equation}
\newpage

Elementary harmonic functions on the manifold have the form
\begin{equation}
h_{lm_{1}m_{2}\cdots m_{N-2}}=h_{l}\left(  w\right)  Y_{lm_{1}m_{2}\cdots
m_{N-2}}\left(  \widehat{r}\right)
\end{equation}
where the radial functions satisfy the equation%
\begin{align}
\frac{l\left(  l+N-2\right)  }{r\left(  w\right)  ^{2}}~h_{l}\left(  w\right)
&  =\frac{1}{r\left(  w\right)  ^{N-1}}\frac{d}{dw}\left(  r\left(  w\right)
^{N-1}\frac{d}{dw}h_{l}\left(  w\right)  \right)  \label{RadialHarmonicEqn}\\
&  =\frac{1}{r\left(  w\right)  }\frac{d}{dw}\left(  r\left(  w\right)
\frac{d}{dw}h_{l}\left(  w\right)  \right)  +\left(  N-2\right)
\frac{r^{\prime}\left(  w\right)  }{r\left(  w\right)  }\frac{d}{dw}%
h_{l}\left(  w\right)
\end{align}
Suppose $h_{l}^{\left(  1\right)  }$ and $h_{l}^{\left(  2\right)  }$ are two
solutions of this second-order differential equation. \ Then by
\href{https://en.wikipedia.org/wiki/Abel's_identity}{Abel's identity} their
\href{https://en.wikipedia.org/wiki/Wronskian}{Wronskian} is%
\begin{equation}
W\left[  h_{l}^{\left(  1\right)  }\left(  w\right)  ,h_{l}^{\left(  2\right)
}\left(  w\right)  \right]  \equiv h_{l}^{\left(  1\right)  }\left(  w\right)
~\overleftrightarrow{\frac{d}{dw}}~h_{l}^{\left(  2\right)  }\left(  w\right)
=\frac{c_{l}}{r^{N-1}\left(  w\right)  }\label{GeneralWronskian}%
\end{equation}
where $c_{l}=r^{N-1}\left(  0\right)  W\left[  h_{l}^{\left(  1\right)
}\left(  0\right)  ,h_{l}^{\left(  2\right)  }\left(  0\right)  \right]  $ is
a constant. \ For the cases of interest, $r\left(  w\right)  $ is monotonic in
$\left\vert w\right\vert $ as well as symmetric under $w\rightarrow-w$ , and
therefore if $h_{l}\left(  w\right)  $ is a solution to
(\ref{RadialHarmonicEqn}), so is $h_{l}\left(  -w\right)  $. \ 

For the rest of the discussion, we assume both $r\left(  w\right)  =r\left(
-w\right)  $ and $r\left(  w\right)  \underset{w\rightarrow\pm\infty}{\sim
}\left\vert w\right\vert $, in which case the asymptotic behavior of the two
radial functions is familiar\footnote{Compare (\ref{AsymptoticHarmonics}) to
the usual radial equation and its solutions on $\mathbb{E}_{N}$,
namely,\ $\frac{1}{r^{N-1}}\frac{d}{dr}\left(  r^{N-1}\frac{d}{dr}h_{l}\left(
r\right)  \right)  =\frac{l\left(  l+N-2\right)  }{r^{2}}h_{l}\left(
r\right)  $, which is solved by $h_{l}\left(  r\right)  =C_{1}r^{l}%
+C_{2}r^{2-N-l}$.} since to leading order (\ref{RadialHarmonicEqn}) reduces to%
\begin{equation}
\frac{1}{\left\vert w\right\vert ^{N-1}}\frac{d}{dw}\left(  \left\vert
w\right\vert ^{N-1}\frac{d}{dw}h_{l}\left(  w\right)  \right)
\underset{r\rightarrow\pm\infty}{\sim}\frac{l\left(  l+N-2\right)
}{\left\vert w\right\vert ^{2}}~h_{l}\left(  w\right)
\label{AsymptoticHarmonics}%
\end{equation}
with simple power law solutions%
\begin{equation}
h_{l}\left(  w\right)  \underset{w\rightarrow\pm\infty}{\sim}\left\vert
w\right\vert ^{l}\text{ \ \ or \ \ \ \ }\frac{1}{\left\vert w\right\vert
^{l+N-2}}%
\end{equation}
For $N>2$ we identify the exact solution that falls off when $w\rightarrow
+\infty$ but does not fall off when $w\rightarrow-\infty$ as $h_{l}^{\left(
+\right)  }$. \ Thus, with a normalization chosen for convenience,%
\begin{equation}
h_{l}^{\left(  +\right)  }\left(  w\right)  \underset{w\rightarrow
+\infty}{\sim}\frac{1}{\left\vert w\right\vert ^{l+N-2}}\ ,\ \ \ h_{l}%
^{\left(  +\right)  }\left(  w\right)  \underset{w\rightarrow-\infty}{\sim
}b_{l}~\left\vert w\right\vert ^{l}%
\end{equation}
where $b_{l}$ is another constant. \ For the cases of interest, for which
$r\left(  w\right)  =r\left(  -w\right)  $, it follows that $h_{l}^{\left(
-\right)  }\left(  w\right)  \equiv h_{l}^{\left(  +\right)  }\left(
-w\right)  $ is an independent exact solution that has the asymptotic behavior%
\begin{equation}
h_{l}^{\left(  -\right)  }\left(  w\right)  \underset{w\rightarrow
-\infty}{\sim}\frac{1}{\left\vert w\right\vert ^{l+N-2}}\ ,\ \ \ h_{l}%
^{\left(  -\right)  }\left(  w\right)  \underset{w\rightarrow+\infty}{\sim
}b_{l}~\left\vert w\right\vert ^{l}%
\end{equation}
Independence follows from the Wronskian%
\begin{align}
W\left[  h_{l}^{\left(  -\right)  }\left(  w\right)  ,h_{l}^{\left(  +\right)
}\left(  w\right)  \right]   &  \equiv h_{l}^{\left(  -\right)  }\left(
w\right)  ~\overleftrightarrow{\frac{d}{dw}}~h_{l}^{\left(  +\right)  }\left(
w\right)  \underset{w\rightarrow+\infty}{\sim}b_{l}\left(  w^{l}~\frac{\left(
-l-N+2\right)  }{w^{l+N-1}}-\frac{1}{w^{l+N-2}}~w^{l-1}l\right) \nonumber\\
&  \underset{w\rightarrow+\infty}{\sim}b_{l}~\frac{\left(  2-N-2l\right)
}{w^{N-1}}%
\end{align}
as expected from the asymptotic form of (\ref{GeneralWronskian}) with
$c_{l}=\left(  2-N-2l\right)  b_{l}$.

For $N>2$ then, a Green function that vanishes as $\left\vert w\right\vert
\rightarrow\infty$ has the form%
\begin{align}
G\left(  w_{1},\widehat{r}_{1};w_{2},\widehat{r}_{2}\right)   &
=\sum_{l,m_{1},m_{2},\cdots,m_{N-2}}\frac{1}{c_{l}}~h_{l}^{\left(  +\right)
}\left(  w_{>}\right)  h_{l}^{\left(  -\right)  }\left(  w_{<}\right)
Y_{lm_{1}m_{2}\cdots m_{N-2}}\left(  \widehat{r_{1}}\right)  Y_{lm_{1}%
m_{2}\cdots m_{N-2}}^{\ast}\left(  \widehat{r_{2}}\right) \nonumber\\
&  =\frac{1}{\left(  2-N\right)  \Omega_{N}}\sum_{l=0}^{\infty}\frac{1}{b_{l}%
}~h_{l}^{\left(  +\right)  }\left(  w_{>}\right)  h_{l}^{\left(  -\right)
}\left(  w_{<}\right)  ~C_{l}^{\left(  \frac{N-2}{2}\right)  }\left(
\cos\theta\right)  \label{GGreenFcn}%
\end{align}
where $w_{>}$ or $w_{<}$ is the $\max$ or $\min$ of $w_{1}$ and $w_{2}$,
respectively. \ When $r\left(  w\right)  =r\left(  -w\right)  $ it follows
that
\begin{equation}
G\left(  w_{1},\widehat{r}_{1};w_{2},\widehat{r}_{2}\right)  =G\left(
w_{2},\widehat{r}_{2};w_{1},\widehat{r}_{1}\right)  \label{Symmetry}%
\end{equation}
and, due to the isotropy of the manifold, the sum over the various $m_{k}$
labeling the hyperspherical harmonics always reduces to a function of just a
single angle $\theta$, namely, $C_{l}^{\left(  \frac{N-2}{2}\right)  }\left(
\cos\theta\right)  $ with $\cos\theta\equiv\widehat{r_{1}}\cdot\widehat{r_{2}%
}$. \ 

Completeness of the $Y_{lm_{1}m_{2}\cdots m_{N-2}}$ on the hypersphere
$S_{N-1}$ and the radial discontinuity given by
\begin{equation}
\lim_{\varepsilon\rightarrow0}\left.  \frac{d}{dw_{1}}\left(  r\left(
w_{1}\right)  ^{N-1}h_{l}^{\left(  +\right)  }\left(  w_{1}\right)  \right)
h_{l}^{\left(  -\right)  }\left(  w_{2}\right)  \right\vert _{w_{1}%
=w_{2}+\varepsilon}-\lim_{\varepsilon\rightarrow0}\left.  h_{l}^{\left(
+\right)  }\left(  w_{2}\right)  \frac{d}{dw_{1}}\left(  r\left(
w_{1}\right)  ^{N-1}h_{l}^{\left(  -\right)  }\left(  w_{1}\right)  \right)
\right\vert _{w_{1}=w_{2}-\varepsilon}=-c_{l}~
\end{equation}
produces the expected invariant Dirac delta in the equation obeyed by $G$,
namely,%
\begin{equation}
\nabla^{2}G\left(  w_{1},\widehat{r}_{1};w_{2},\widehat{r}_{2}\right)
=-\frac{1}{r\left(  w\right)  ^{N-1}}~\delta\left(  w_{1}-w_{2}\right)
~\delta^{N-1}\left(  \widehat{r_{1}}-\widehat{r_{2}}\right)
\label{GreenFunctionEquation}%
\end{equation}

If either $w_{>}\rightarrow+\infty$ or $w_{<}\rightarrow-\infty$, the
individual terms vanish in the sum (\ref{GGreenFcn}), so that the Green
function satisfies the homogeneous boundary condition $G=0$, assuming
convergence of the sum. \ However, when $w_{2}=0$ the value of $G$ is not so
obvious. \ Nevertheless, a simple linear combination can be selected to
construct a Green function that vanishes at $w_{2}=0$, namely,%
\begin{equation}
G_{o}\left(  w_{1},\widehat{r}_{1};w_{2},\widehat{r}_{2}\right)  =G\left(
w_{1},\widehat{r}_{1};w_{2},\widehat{r}_{2}\right)  -G\left(  w_{1}%
,\widehat{r}_{1};-w_{2},\widehat{r}_{2}\right)  \label{GoGreenFcn}%
\end{equation}
Note that $G_{o}$ is manifestly an odd function of $w_{2}$ and is also an odd
function of $w_{1}$ as a consequence of (\ref{Symmetry}).

This $G_{o}$ has a simple interpretation as the potential at $\left(
w_{1},\widehat{r}_{1}\right)  $ due to a point charge source at $\left(
w_{2},\widehat{r}_{2}\right)  $ and a negative image charge of that point
source at $\left(  -w_{2},\widehat{r}_{2}\right)  $. \ The source and image
charges are therefore of \emph{equal} magnitude but opposite sign, and are
positioned symmetrically but on \emph{opposite} branches of the manifold.
\ So, if both $w_{1}$ and $w_{2}$ are restricted to one branch of the
manifold, (\ref{GreenFunctionEquation}) will still hold on that branch, since
the image will produce an additional Dirac delta only on the other branch.
\ Therefore $G_{o}$ is an appropriate Green function to solve $\nabla^{2}%
\Phi=-\rho$ on the upper branch of the manifold with the condition $\Phi=0$ on
the inner boundary of that branch, i.e. on the hypersphere at $w=0$.

The radial equation (\ref{RadialHarmonicEqn}) is perhaps more transparent
after changing variable to
\begin{equation}
u=\int_{0}^{w}\frac{d\varpi}{r\left(  \varpi\right)  } \label{uVariable}%
\end{equation}
with implicit inverse $w\left(  u\right)  $. \ Note that $\operatorname*{sgn}%
u=\operatorname*{sgn}w$, and if $r\left(  w\right)  \underset{w\rightarrow
\pm\infty}{\sim}\left\vert w\right\vert $ then $u\underset{w\rightarrow
\pm\infty}{\rightarrow}\pm\infty$. \ With this variable change, the radial
parts of harmonic functions satisfy the equation%
\begin{equation}
\frac{d^{2}}{du^{2}}h_{l}+\left(  N-2\right)  r^{\prime}\left(  w\left(
u\right)  \right)  \frac{d}{du}h_{l}=l\left(  l+N-2\right)  h_{l}%
\end{equation}
This may be cast into
\href{https://en.wikipedia.org/wiki/Sturm-Liouville_theory}{Sturm-Liouville}
form on the interval $-1\leq t\leq+1$\ upon changing variables to $t=\tanh u$,
and multiplying by the integrating factor \
\begin{equation}
\mu\left(  t\right)  =\exp\left(  \int_{0}^{t}\frac{\left(  N-2\right)
r^{\prime}\left(  w\left(  \operatorname{arctanh}\tau\right)  \right)
}{\left(  1-\tau^{2}\right)  }d\tau\right)
\end{equation}
The result is%
\begin{equation}
\frac{d}{dt}\left(  \mu\left(  t\right)  \left(  1-t^{2}\right)  \frac{d}%
{dt}h_{l}\right)  =\frac{l\left(  l+N-2\right)  }{1-t^{2}}~\mu\left(
t\right)  h_{l}%
\end{equation}
So written, the reader can easily
\href{https://search.yahoo.com/search;_ylt=AwrE19d45RNbb1AAJ9lXNyoA;_ylc=X1MDMjc2NjY3OQRfcgMyBGZyA3lmcC10BGdwcmlkA2htcEFhNXN6VC5xbmY3NGNHWlNlS0EEbl9yc2x0AzAEbl9zdWdnAzIEb3JpZ2luA3NlYXJjaC55YWhvby5jb20EcG9zAzAEcHFzdHIDBHBxc3RybAMwBHFzdHJsAzM2BHF1ZXJ5A3N0dXJtJTIwbGlvdXZpbGxlJTIwZ3JlZW4lMjBmdW5jdGlvbgR0X3N0bXADMTUyODAzMDU5Ng--?p=sturm+liouville+green+function&fr2=sb-top&fr=yfp-t&fp=1}{find
in the mathematical literature} several detailed discussions of Green
functions for this equation.

\section{The Ellis wormhole in $N$ dimensions}

The so-called Ellis wormhole \cite{Ellis} is defined by
(\ref{GeneralIsotropicWormhole}) with%
\begin{equation}
r\left(  w\right)  =\sqrt{R^{2}+w^{2}}%
\end{equation}
or $w=\pm\sqrt{r^{2}-R^{2}}$ on the upper\ and lower\ branches of the
wormhole, respectively. \ In this case%
\begin{align}
u\left(  w\right)   &  =\int_{0}^{w}\frac{dv}{\sqrt{R^{2}+v^{2}}%
}=\operatorname{arcsinh}\left(  \frac{w}{R}\right)  =\ln\left(  \frac{w}%
{R}+\sqrt{1+\frac{w^{2}}{R^{2}}}\right) \\
w\left(  u\right)   &  =R\sinh u\ ,\ \ \ r\left(  w\left(  u\right)  \right)
=R\cosh u\ ,\ \ \ r^{\prime}\left(  w\left(  u\right)  \right)  =\frac
{w\left(  u\right)  }{\sqrt{R^{2}+w^{2}\left(  u\right)  }}=\tanh u
\end{align}
and therefore%
\begin{equation}
\frac{d^{2}}{du^{2}}h_{l}+\left(  N-2\right)  \left(  \tanh u\right)  \frac
{d}{du}h_{l}=l\left(  l+N-2\right)  h_{l} \label{EllisDiffEqnND}%
\end{equation}
This has exact solutions%
\begin{equation}
h_{l}^{\left(  1\right)  }\left(  u\right)  =\frac{\left(  1+\tanh u\right)
^{\frac{1}{2}\left(  N+l-2\right)  }}{\left(  1-\tanh u\right)  ^{\frac{1}%
{2}l}}\left.  _{2}F_{1}\right.  \left(  2-\frac{1}{2}N,\frac{1}{2}%
N-1;2-\frac{1}{2}N-l;\frac{1-\tanh u}{2}\right)  \label{1stEllisRadial}%
\end{equation}%
\begin{equation}
h_{l}^{\left(  2\right)  }\left(  u\right)  =\frac{\left(  1-\tanh u\right)
^{\frac{1}{2}\left(  N+l-2\right)  }}{\left(  1+\tanh u\right)  ^{\frac{1}%
{2}l}}\left.  _{2}F_{1}\right.  \left(  2-\frac{1}{2}N,\frac{1}{2}N-1;\frac
{1}{2}N+l;\frac{1-\tanh u}{2}\right)  \label{2ndEllisRadial}%
\end{equation}
when written in terms of Gauss hypergeometric functions,%
\begin{equation}
\left.  _{2}F_{1}\right.  \left(  a,b;c;z\right)  =\sum_{k=0}^{\infty}%
\frac{\left(  a\right)  _{k}\left(  b\right)  _{k}}{\left(  c\right)  _{k}%
}\frac{z^{k}}{k!}%
\end{equation}
where $\left(  a\right)  _{k}=\Gamma\left(  a+k\right)  /\Gamma\left(
a\right)  $, etc. \ It follows that $h_{l}^{\left(  1\right)  }$ is
well-behaved as $u\rightarrow-\infty$ for all even $N>2$, but \emph{not} for
odd $N$, where it is necessary to take a linear combination of $h_{l}^{\left(
1\right)  }$ and $h_{l}^{\left(  2\right)  }$ to find good behavior. \ On the
other hand, $h_{l}^{\left(  2\right)  }$ is well-behaved as $u\rightarrow
+\infty$ for all $N>2$. \ (For $N=2$, see \cite{CurtrightEtAl}.)

Evidently the simplest case beyond two dimensions is $N=4$, for which the
general solution of (\ref{EllisDiffEqnND}) is given by%
\begin{equation}
h_{l}\left(  u\right)  =\frac{1}{\cosh u}~\left(  c_{1}\exp\left[  \left(
l+1\right)  u\right]  +c_{2}\exp\left[  -\left(  l+1\right)  u\right]
\right)
\end{equation}
for any constants $c_{1}$ and $c_{2}$. \ That is to say,
\begin{gather}
h_{l}^{\left(  1,2\right)  }\left(  u\right)  =\frac{1}{\cosh u}~\exp\left[
\pm\left(  l+1\right)  u\right]  \ ,\ \ \ h_{l}^{\left(  1\right)  }\left(
-\infty\right)  =0\ ,\ \ \ h_{l}^{\left(  2\right)  }\left(  +\infty\right)
=0\\
W\left[  h_{l}^{\left(  1\right)  }\left(  u\right)  ,h_{l}^{\left(  2\right)
}\left(  u\right)  \right]  \equiv h_{l}^{\left(  1\right)  }\left(  u\right)
~\overleftrightarrow{\frac{d}{du}}~h_{l}^{\left(  2\right)  }\left(  u\right)
=-\frac{2\left(  l+1\right)  }{\cosh^{2}u}%
\end{gather}
Or, in terms of the variable $t=\tanh u=\frac{w}{\sqrt{R^{2}+w^{2}}}$, for
$N=4$\ the two independent solutions (\ref{1stEllisRadial}) and
(\ref{2ndEllisRadial}) become%
\begin{gather}
h_{l}^{\left(  1,2\right)  }\left(  t\right)  =\frac{\left(  1\pm t\right)
^{l+1}}{\left(  1-t^{2}\right)  ^{l/2}}\ ,\ \ \ h_{l}^{\left(  1\right)
}\left(  -1\right)  =0\ ,\ \ \ h_{l}^{\left(  2\right)  }\left(  +1\right)
=0\label{N=4EllisHarmonics}\\
W\left[  h_{l}^{\left(  1\right)  }\left(  t\right)  ,h_{l}^{\left(  2\right)
}\left(  t\right)  \right]  \equiv h_{l}^{\left(  1\right)  }\left(  t\right)
~\overleftrightarrow{\frac{d}{dt}}~h_{l}^{\left(  2\right)  }\left(  t\right)
=-2\left(  l+1\right)
\end{gather}

In terms of these harmonic functions for $N=4$, the Green function
(\ref{GGreenFcn}) is%
\begin{align}
&  G\left(  w_{1},\widehat{r}_{1};w_{2},\widehat{r}_{2}\right)
\label{N=4EllisGreenSeriesSummed}\\
&  =\frac{1}{4\pi^{2}\sqrt{\left(  R^{2}+w_{>}^{2}\right)  \left(  R^{2}%
+w_{<}^{2}\right)  }}\sum_{l=0}^{\infty}~\left(  \sqrt{\frac{\left(
w_{<}+\sqrt{R^{2}+w_{<}^{2}}\right)  \left(  w_{>}-\sqrt{R^{2}+w_{>}^{2}%
}\right)  }{\left(  w_{>}+\sqrt{R^{2}+w_{>}^{2}}\right)  \left(  w_{<}%
-\sqrt{R^{2}+w_{<}^{2}}\right)  }}\right)  ^{l+1}~C_{l}^{\left(  1\right)
}\left(  \widehat{r_{1}}\cdot\widehat{r_{2}}\right) \nonumber\\
&  =\frac{1}{4\pi^{2}\sqrt{\left(  R^{2}+w_{>}^{2}\right)  \left(  R^{2}%
+w_{<}^{2}\right)  }}\frac{1}{\sqrt{\frac{\left(  w_{>}+\sqrt{R^{2}+w_{>}^{2}%
}\right)  \left(  w_{<}-\sqrt{R^{2}+w_{<}^{2}}\right)  }{\left(  w_{<}%
+\sqrt{R^{2}+w_{<}^{2}}\right)  \left(  w_{>}-\sqrt{R^{2}+w_{>}^{2}}\right)
}}+\sqrt{\frac{\left(  w_{<}+\sqrt{R^{2}+w_{<}^{2}}\right)  \left(
w_{>}-\sqrt{R^{2}+w_{>}^{2}}\right)  }{\left(  w_{>}+\sqrt{R^{2}+w_{>}^{2}%
}\right)  \left(  w_{<}-\sqrt{R^{2}+w_{<}^{2}}\right)  }}-2\widehat{r_{1}%
}\cdot\widehat{r_{2}}}\nonumber
\end{align}
In the last line we have used the sum%
\begin{equation}
\sum_{l=0}^{\infty}x^{l}~C_{l}^{\left(  1\right)  }\left(  \cos\theta\right)
=\frac{1}{1+x^{2}-2x\cos\theta}%
\end{equation}
This result for the $N=4$\ Green function is more compactly written as%
\begin{equation}
G\left(  w_{1},\widehat{r}_{1};w_{2},\widehat{r}_{2}\right)  =\frac{1}%
{4\pi^{2}r_{1}r_{2}\left(  \sqrt{\frac{\left(  r_{1}+w_{1}\right)  \left(
r_{2}-w_{2}\right)  }{\left(  r_{2}+w_{2}\right)  \left(  r_{1}-w_{1}\right)
}}+\sqrt{\frac{\left(  r_{2}+w_{2}\right)  \left(  r_{1}-w_{1}\right)
}{\left(  r_{1}+w_{1}\right)  \left(  r_{2}-w_{2}\right)  }}-2\widehat{r_{1}%
}\cdot\widehat{r_{2}}\right)  } \label{N=4EllisGreen}%
\end{equation}
where $r_{1}\equiv\sqrt{R^{2}+w_{1}^{2}}$ and $r_{2}\equiv\sqrt{R^{2}%
+w_{2}^{2}}$. \ Note that $G\left(  w_{1},\widehat{r}_{1};w_{2},\widehat{r}%
_{2}\right)  =G\left(  w_{2},\widehat{r}_{2};w_{1},\widehat{r}_{1}\right)  $.
\ The $N=4$ Green function for the grounded wormhole is then defined as in
(\ref{GoGreenFcn}), so that $G_{o}\left(  w_{1},\widehat{r}_{1};w_{2}%
,\widehat{r}_{2}\right)  =-G_{o}\left(  -w_{1},\widehat{r}_{1};w_{2}%
,\widehat{r}_{2}\right)  $. \ 

Asymptotically, with both points on the upper branch of the manifold,%
\begin{equation}
G\left(  w_{1},\widehat{r}_{1};w_{2},\widehat{r}_{2}\right)  \underset{w_{1}%
,w_{2}\gg R}{\sim}\frac{1}{4\pi^{2}\left\vert \overrightarrow{r_{1}%
}-\overrightarrow{r_{2}}\right\vert ^{2}}+O\left(  \frac{R}{r_{1,2}^{3}%
}\right)  \label{N=4Asymptotics}%
\end{equation}
As should be expected, the leading term here is in agreement with
(\ref{EuclideanGreen}) for $N=4$. \ 

A contour plot of $G$ versus $w_{1}$ and $\theta\equiv\arccos\left(
\widehat{r_{1}}\cdot\widehat{r_{2}}\right)  $ is shown for the $N=4$ Ellis
wormhole in Figure 10, with $R=1$ and unit source at $\left(  w_{2}%
,\widehat{r_{2}}\right)  =\left(  1,\widehat{r_{2}}\right)  $. \ A similar
plot of $G_{o}$ versus $w_{1}$ and $\theta$ is shown in Figure 11. \ In
addition to the unit source at $\left(  w_{2},\widehat{r_{2}}\right)  =\left(
1,\widehat{r_{2}}\right)  $, $G_{o}$\ incorporates a negative image of that
source at $\left(  w_{2},\widehat{r_{2}}\right)  =\left(  -1,\widehat{r_{2}%
}\right)  $.

\section{The $p$-norm wormholes}

Consider next a continuous deformation of the Ellis wormhole. \ Define a class
of \textquotedblleft$p$-norm radial functions\textquotedblright\ with $R$ a
constant radius, $p\geq1$ a real number, and
\begin{equation}
r\left(  w\right)  =\left(  R^{p}+\left\vert w\right\vert ^{p}\right)  ^{1/p}
\label{pNormRadius}%
\end{equation}
Note that indeed $r\left(  w\right)  \underset{\left\vert w\right\vert \gg
R}{\sim}\left\vert w\right\vert $, so the asymptotic behavior of harmonic
functions on these manifolds falls within the scope of the discussion
following (\ref{AsymptoticHarmonics}). \ Also note the case $p=2$ is the Ellis
wormhole. \ Equatorial slices of the manifolds defined by (\ref{pNormRadius})
and (\ref{GeneralIsotropicWormhole}) for various values of $p$ are shown in
the Figures\ as 3D embeddings of 2D surfaces, to obtain curved surfaces of
revolution about the $z$-axis. \ 

As $p\rightarrow1$, or else as $p\rightarrow\infty$, this class of manifolds
continuously interpolates between the Ellis wormhole, with its smoothly curved
bridge, and two flattened copies of $\mathbb{E}_{N}$ each of which is missing
an $N$-ball $\mathbb{B}_{N}$ of radius $R$. \ That is to say, varying $p$ away
from $p=2$ interpolates between the Ellis wormhole and a pair of
$\mathbb{E}_{N}-\mathbb{B}_{N}\left(  R\right)  $ manifolds, either as
$p\rightarrow1$ or else as $p\rightarrow\infty$. \ Nonetheless, the two copies
of $\mathbb{E}_{N}-\mathbb{B}_{N}\left(  R\right)  $ are joined together,
either by a single $S_{N-1}$ of radius $R$, as $p\rightarrow1$, or by a tube
composed of such $S_{N-1}$s, as $p\rightarrow\infty$. \ For the first of these
limits, we say the two copies of $\mathbb{E}_{N}-\mathbb{B}_{N}\left(
R\right)  $ are \textquotedblleft creased\textquotedblright\ together on a
single hypersphere of radius $R$, that hypersphere providing an open
\textquotedblleft doorway\textquotedblright\ to go from one copy of
$\mathbb{E}_{N}-$ $\mathbb{B}_{N}\left(  R\right)  $ to the other.

For $G$ and $G_{o}$, in this short section on generic $p$-norm manifolds we
are content to refer to the previous general discussion of the Green
functions. \ To supplement that discussion, we only point out that the
variable $u$ defined in (\ref{uVariable}) is explicitly given for the $p$-norm
radial functions by Gauss hypergeometric functions, namely,%
\begin{equation}
u\left(  w\right)  =\int_{0}^{w}\frac{1}{\left(  R^{p}+\left(  \varpi
^{2}\right)  ^{p/2}\right)  ^{1/p}}~d\varpi=\frac{w}{R}\left.  _{2}%
F_{1}\right.  \left(  \frac{1}{p},\frac{1}{p};1+\frac{1}{p};-\left(
\frac{w^{2}}{R^{2}}\right)  ^{\frac{1}{2}p}\right)
\end{equation}

\section{The squashed wormhole and its Green functions}

Flattening the two branches of the wormhole, to obtain two independent copies
of $\mathbb{E}_{N}-\mathbb{B}_{N}\left(  R\right)  $ joined together on a
common hypersphere, is achieved by taking the $p=1$ Manhattan
norm.\footnote{In the opposite extreme, when $p\rightarrow\infty$ an
equatorial slice of the $p$-norm wormhole becomes the right-circular
cylindrical tube of \cite{JamesEtAl}, p 488, Eqn (3).}\
\begin{equation}
r\left(  w\right)  =R+\left\vert w\right\vert
\end{equation}
Then so long as $w\neq0$ (\ref{RadialHarmonicEqn}) is solved by familiar
functions,%
\begin{equation}
h_{l}\left(  w\right)  =r\left(  w\right)  ^{l}\text{ \ \ or \ \ \ \ }r\left(
w\right)  ^{2-N-l}%
\end{equation}
It suffices to consider two situations for the source and field point
locations, either with $w_{1}$ and $w_{2}$ on the same branch of the manifold,
or with $w_{1}$ and $w_{2}$ on opposite branches. \ 

Suppose $w_{>}$ is always on the upper branch, say. \ Then (\ref{GGreenFcn})
will have different forms for the two possible source and field point
locations.
\begin{align}
G\left(  w_{1},\widehat{r}_{1};w_{2},\widehat{r}_{2}\right)   &  =k_{N}%
\sum_{l=0}^{\infty}\frac{\left(  r\left(  w_{<}\right)  \right)  ^{l}}{\left(
r\left(  w_{>}\right)  \right)  ^{l+N-2}}~C_{l}^{\left(  \frac{N-2}{2}\right)
}\left(  \widehat{r_{1}}\cdot\widehat{r_{2}}\right)  \text{ \ \ \ \ if both
}w_{>}>0\text{ \& }w_{<}>0\label{GUpUp}\\
&  =k_{N}\sum_{l=0}^{\infty}\frac{\left(  R\right)  ^{2l+N-2}}{\left(
r\left(  w_{>}\right)  r\left(  w_{<}\right)  \right)  ^{l+N-2}}%
~C_{l}^{\left(  \frac{N-2}{2}\right)  }\left(  \widehat{r_{1}}\cdot
\widehat{r_{2}}\right)  \text{ \ \ \ \ if }w_{>}>0\text{ but}\ w_{<}<0
\label{GUpLow}%
\end{align}
The two forms are chosen so that $G\rightarrow0$ as $w_{>}\rightarrow+\infty$
or as $w_{<}\rightarrow-\infty$, and so that $G$ is continuous as
$w_{<}\rightarrow0$, i.e. at $r\left(  w_{<}\right)  =R$. \ 

For the first situation with\ $w_{1}>0$ and $w_{2}>0$, in light of
(\ref{GegenbauerExpansion}) the sum in (\ref{GUpUp}) gives%
\begin{align}
\left.  G\left(  w_{1},\widehat{r}_{1};w_{2},\widehat{r}_{2}\right)
\right\vert _{\substack{w_{1}>0\\w_{2}>0}}  &  =\frac{k_{N}}{\left(
r^{2}\left(  w_{1}\right)  +r^{2}\left(  w_{2}\right)  -2r\left(
w_{1}\right)  r\left(  w_{2}\right)  \widehat{r_{1}}\cdot\widehat{r_{2}%
}\right)  ^{\frac{N-2}{2}}}\label{Gflat++}\\
&  =\frac{k_{N}}{\left\vert \overrightarrow{r_{1}}-\overrightarrow{r_{2}%
}\right\vert ^{N-2}}%
\end{align}
where in the last expression we have identified $\overrightarrow{r_{1}%
}=\left(  R+\left\vert w_{1}\right\vert \right)  \widehat{r}_{1}$ and
$\overrightarrow{r_{2}}=\left(  R+\left\vert w_{2}\right\vert \right)
\widehat{r}_{2}$ to obtain a symmetrical form that can also be used if both
$w_{1}<0$ and $w_{2}<0$. \ 

For the second situation with $w_{1}>0$ and $w_{2}<0$, the sum in
(\ref{GUpLow}) gives
\begin{align}
\left.  G\left(  w_{1},\widehat{r}_{1};w_{2},\widehat{r}_{2}\right)
\right\vert _{\substack{w_{1}>0\\w_{2}<0}}  &  =\frac{R^{N-2}k_{N}}{\left(
r^{2}\left(  w_{1}\right)  r^{2}\left(  w_{2}\right)  +R^{4}-2R^{2}r\left(
w_{1}\right)  r\left(  w_{2}\right)  \widehat{r_{1}}\cdot\widehat{r_{2}%
}\right)  ^{\frac{N-2}{2}}}\label{Gflat+-}\\
&  =\frac{R^{N-2}k_{N}}{\left(  r_{1}^{2}r_{2}^{2}+R^{4}-2R^{2}%
\overrightarrow{r_{1}}\cdot\overrightarrow{r_{2}}\right)  ^{\frac{N-2}{2}}}%
\end{align}
where in the last expression we have again identified $\overrightarrow{r_{1}%
}=\left(  R+\left\vert w_{1}\right\vert \right)  \widehat{r}_{1}$ and
$\overrightarrow{r_{2}}=\left(  R+\left\vert w_{2}\right\vert \right)
\widehat{r}_{2}$ to obtain a symmetrical form that can also be used if
$w_{1}<0$ and $w_{2}>0$. \ 

So, as might have been anticipated, $G\left(  w_{1},\widehat{r}_{1}%
;w_{2},\widehat{r}_{2}\right)  =G\left(  w_{2},\widehat{r}_{2};w_{1}%
,\widehat{r}_{1}\right)  $ in all situations. \ Taken together, (\ref{Gflat++}%
) and (\ref{Gflat+-}) give\ $G$ as a solution to (\ref{GreenFunctionEquation})
for any field and source point locations on the squashed manifold. \ A contour
plot of $G$ versus $w_{1}$ and $\theta\equiv\arccos\left(  \widehat{r_{1}%
}\cdot\widehat{r_{2}}\right)  $ is shown for the squashed wormhole in Figure
12, with $N=4$, $R=1$, and unit source at $\left(  w_{2},\widehat{r_{2}%
}\right)  =\left(  1,\widehat{r_{2}}\right)  $. \ 

Once again a simple linear combination can be taken to construct a Green
function that vanishes at $w_{2}=0$, as in (\ref{GoGreenFcn}). \ Given that
$r\left(  w\right)  =r\left(  -w\right)  $, when both $w_{1}>0$ and $w_{2}>0$
this grounded Green function for the squashed wormhole is explicitly
\begin{equation}
\left.  G_{o}\left(  w_{1},\widehat{r}_{1};w_{2},\widehat{r}_{2}\right)
\right\vert _{\substack{w_{1}>0\\w_{2}>0}}=\frac{k_{N}}{\left\vert
\overrightarrow{r_{1}}-\overrightarrow{r_{2}}\right\vert ^{N-2}}-\frac
{R^{N-2}k_{N}}{\left(  r_{1}^{2}r_{2}^{2}+R^{4}-2R^{2}\overrightarrow{r_{1}%
}\cdot\overrightarrow{r_{2}}\right)  ^{\frac{N-2}{2}}}
\label{GreenGroundedSquashedUpper}%
\end{equation}
Recall that $G_{o}$ is not only manifestly an odd function of $w_{2}$ but is
also an odd function of $w_{1}$ as a consequence of (\ref{GoGreenFcn}) and the
symmetry $G\left(  w_{1},\widehat{r}_{1};w_{2},\widehat{r}_{2}\right)
=G\left(  w_{2},\widehat{r}_{2};w_{1},\widehat{r}_{1}\right)  $. \ Therefore,
when the field point is on the lower branch of the squashed wormhole with the
source on the upper branch,
\begin{equation}
\left.  G_{o}\left(  w_{1},\widehat{r}_{1};w_{2},\widehat{r}_{2}\right)
\right\vert _{\substack{w_{1}<0\\w_{2}>0}}=-\frac{k_{N}}{\left\vert
\overrightarrow{r_{1}}-\overrightarrow{r_{2}}\right\vert ^{N-2}}+\frac
{R^{N-2}k_{N}}{\left(  r_{1}^{2}r_{2}^{2}+R^{4}-2R^{2}\overrightarrow{r_{1}%
}\cdot\overrightarrow{r_{2}}\right)  ^{\frac{N-2}{2}}}
\label{GreenGroundedSquashedLower}%
\end{equation}
Again, for emphasis, this $G_{o}$ has the interpretation as the potential at
$\left(  w_{1},\widehat{r}_{1}\right)  $ due to a point charge source at
$\left(  w_{2},\widehat{r}_{2}\right)  $ and a negative image charge of that
point source at $\left(  -w_{2},\widehat{r}_{2}\right)  $. \ The source and
image charges are therefore of \emph{equal} magnitude but opposite sign, and
are positioned in an \emph{natural} way on \emph{opposite} branches of the
manifold. \ 

At the risk of being repetitive, if both $w_{1}$ and $w_{2}$ are restricted to
one branch of the manifold, (\ref{GreenFunctionEquation}) will still hold on
that branch, since the image will produce an additional Dirac delta only on
the other branch. \ Therefore $G_{o}$ is an appropriate Green function to
solve $\nabla^{2}\Phi=-\rho$ on the upper branch of the manifold with the
condition $\Phi=0$ on the inner boundary of that branch, i.e. on the
hypersphere at $w=0$.

When both field and source points are on the upper branch of the wormhole,
this result for $G_{o}$ is exactly the usual result for the Green function in
the region exterior to a grounded hypersphere, as obtained by considering only
\emph{one copy} of $\mathbb{E}_{N}$ and putting a unit source at
$\overrightarrow{r_{2}}$ outside the hypersphere along with a negative image
source of \emph{reduced strength} $-\left(  R/r_{2}\right)  ^{N-2}$ at an
\textquotedblleft inversion point\textquotedblright\ $\overrightarrow{r}%
_{\text{image}}=\frac{R^{2}}{r_{2}^{2}}~\overrightarrow{r_{2}}$ that
lies\ \emph{inside} the hypersphere, namely,%
\begin{equation}
G_{0}\left(  \overrightarrow{r_{1}};\overrightarrow{r_{2}}\right)
=\frac{k_{N}}{\left\vert \overrightarrow{r_{1}}-\overrightarrow{r_{2}%
}\right\vert ^{N-2}}-\frac{R^{N-2}}{r_{2}^{N-2}}\frac{k_{N}}{\left\vert
\overrightarrow{r_{1}}-\frac{R^{2}}{r_{2}^{2}}~\overrightarrow{r_{2}%
}\right\vert ^{N-2}}=G_{0}\left(  \overrightarrow{r_{2}};\overrightarrow{r_{1}%
}\right)  \label{StandardGo}%
\end{equation}
Thus, either image procedure gives the same $G_{o}$ Green function when
restricted to this single copy of $\mathbb{E}_{N}-\mathbb{B}_{N}\left(
R\right)  $. In particular, for $N=4$, on the upper branch of the wormhole,%
\begin{equation}
G_{0}\left(  \overrightarrow{r_{1}},\overrightarrow{r_{2}}\right)  =\frac
{1}{4\pi^{2}\left\vert \overrightarrow{r_{1}}-\overrightarrow{r_{2}%
}\right\vert ^{2}}-\frac{R^{2}}{r_{2}^{2}}\frac{1}{4\pi^{2}\left\vert
\overrightarrow{r_{1}}-\frac{R^{2}}{r_{2}^{2}}~\overrightarrow{r_{2}%
}\right\vert ^{2}}=\frac{1}{4\pi^{2}\left(  r_{1}^{2}+r_{2}^{2}%
-2\overrightarrow{r_{1}}\cdot\overrightarrow{r_{2}}\right)  }-\frac{R^{2}%
}{4\pi^{2}\left(  r_{1}^{2}r_{2}^{2}+R^{4}-2R^{2}\overrightarrow{r_{1}}%
\cdot\overrightarrow{r_{2}}\right)  }%
\end{equation}
A contour plot of $G_{o}$ for the $N=4$ squashed wormhole is shown in Figure
13, for $\overrightarrow{r_{1}}$ on both upper and lower branches, with a unit
source\ on the upper branch at $\left(  w_{2},\theta_{2}\right)  =\left(
1,0\right)  $ and its negative image on the lower branch at $\left(
-w_{2},\theta_{2}\right)  =\left(  -1,0\right)  $.

\section{Relating image charge distributions by inversion}

Coordinate inversion for a single copy of $\mathbb{E}_{N}$\ maps the interior
of the hypersphere to the exterior, and vice versa, and therefore inversion
should be expected to relate the standard image method, where the so-called
\href{https://en.wikipedia.org/wiki/William_Thomson,_1st_Baron_Kelvin}{Kelvin}
image is placed inside the hypersphere, to the Sommerfeld method for the
squashed wormhole. \ Indeed, inversion of the source position is the technique
that is normally invoked to locate Kelvin images for grounded hypersphere
Green functions on $\mathbb{E}_{N}$. \ 

The inversion mapping is defined by%
\begin{equation}
\overrightarrow{\mathfrak{r}}=\frac{R^{2}}{r^{2}}~\overrightarrow{r}%
\end{equation}
Radial distances change under the inversion, $\mathfrak{r}=R^{2}/r$, but
angles do not, $\widehat{\mathfrak{r}}=\widehat{r}$. \ Under an inversion the
Laplacian does \emph{not} transform into a geometric factor multiplying just
the Laplacian, \emph{except when }$N=2$. \ In other dimensions, the Laplacian
mixes with the \emph{scale operator} under an inversion, as follows.%
\begin{equation}
\nabla_{r}^{2}=\left(  \frac{\mathfrak{r}^{2}}{R^{2}}\right)  ^{2}\left(
\nabla_{\mathfrak{r}}^{2}+\frac{2\left(  2-N\right)  }{\mathfrak{r}^{2}%
}~D_{\mathfrak{r}}\right)  \ ,\ \ \ \ \ D_{\mathfrak{r}}%
=\overrightarrow{\mathfrak{r}}\cdot\overrightarrow{\nabla}_{\mathfrak{r}}
\label{InversionOnLaplacian}%
\end{equation}
This statement may be understood by considering harmonic functions, upon
noting that\ under inversions $r$ effectively becomes $1/r$, and only when
$N=2$ do both $r^{l}$ and $r^{-l}$ appear as factors in harmonic functions.
\ For other $N$ the factors are $r^{l}$ and $r^{2-N-l}$.

Moreover, $G$ itself is \emph{not} invariant under the inversion, \emph{except
when }$N=2$. \ This is obvious on dimensional grounds, since $G\left(
\overrightarrow{r},0\right)  \propto1/r^{N-2}\underset{\text{inversion}%
}{\longrightarrow}\mathfrak{r}^{N-2}/R^{2N-4}\propto\left(  \mathfrak{r}%
^{2N-4}/R^{2N-4}\right)  G\left(  \overrightarrow{\mathfrak{r}},0\right)  $.
\ More precisely, in $N$ spatial dimensions,%
\begin{equation}
G\left(  \overrightarrow{r_{1}};\overrightarrow{r_{2}}\right)
\underset{\text{inversion}}{\longrightarrow}\frac{\mathfrak{r}_{1}%
^{N-2}~\mathfrak{r}_{2}^{N-2}}{R^{2N-4}}~G\left(  \overrightarrow{\mathfrak{r}%
_{1}};\overrightarrow{\mathfrak{r}_{2}}\right)
\end{equation}
Note the symmetry under $\overrightarrow{r_{1}}\leftrightarrow
\overrightarrow{r_{2}}$\ is maintained under $\overrightarrow{\mathfrak{r}%
_{1}}\leftrightarrow\overrightarrow{\mathfrak{r}_{2}}$. \ The complete
transformation of the differential equation for the Green function in $N$
dimensions is%
\begin{align}
\nabla_{r_{1}}^{2}G\left(  \overrightarrow{r_{1}};\overrightarrow{r_{2}%
}\right)   &  =-\frac{1}{\sqrt{g}}~\delta^{N}\left(  \overrightarrow{r_{1}%
}-\overrightarrow{r_{2}}\right)  \underset{\text{inversion}}{\longrightarrow
}\\
\left(  \frac{\mathfrak{r}_{1}^{2}}{R^{2}}\right)  ^{2}\left(  \nabla
_{\mathfrak{r}_{1}}^{2}+\frac{2\left(  2-N\right)  }{\mathfrak{r}_{1}^{2}%
}~D_{\mathfrak{r}_{1}}\right)  \left(  \frac{\mathfrak{r}_{1}^{N-2}%
~\mathfrak{r}_{2}^{N-2}}{R^{2N-4}}~G\left(  \overrightarrow{\mathfrak{r}_{1}%
};\overrightarrow{\mathfrak{r}_{2}}\right)  \right)   &  =-\frac
{\mathfrak{r}_{1}^{N+1}}{R^{2N}}~\delta\left(  \mathfrak{r}_{1}-\mathfrak{r}%
_{2}\right)  ~\delta^{N-1}\left(  \widehat{\mathfrak{r}}_{1}%
-\widehat{\mathfrak{r}}_{2}\right)
\end{align}
Again the $N=2$ case is especially simple. \ For $N=2$, up to a common factor,
the equation is unchanged in form by the inversion.

In view of these results, it is not difficult to map only the lower branch of
the squashed wormhole into the interior of the hypersphere while leaving the
upper branch unchanged, thereby obtaining a single copy of $\mathbb{E}_{N}$
that includes both the exterior and the interior of the hypersphere. \ In the
course of this inversion, the image charge is moved to its more conventional
position within the hypersphere. \ We leave the details as an exercise for the reader.

\section{On the grounded conducting disk in 3D}

\href{https://en.wikipedia.org/wiki/E._W._Hobson}{Hobson} used Sommerfeld's
method and a clever coordinate choice to find the Green function for an
equipotential circular disk in three Euclidean dimensions \cite{Hobson}. \ In
this approach, the disk serves as a doorway between two copies of
$\mathbb{E}_{3}$, with the unit source and field point located in one copy of
$\mathbb{E}_{3}$,\ and an equal strength, negative image of the source
obviously placed in the same position as the source except in the second copy
of $\mathbb{E}_{3}$. \ Nearly forty years later, Waldmann (a student of
Sommerfeld) offered another solution to this problem \cite{Waldmann} by
mapping the half-plane to a finite radius disk and then transforming
Sommerfeld's 1897 result for the grounded half-plane Green function. \ Another
thirty-four years after that, Davis and Reitz independently solved the same
problem, again using Sommerfeld's method but with an emphasis on the use of
complex analysis to construct directly the Green function on the two copies of
$\mathbb{E}_{3}$ \cite{DavisReitz}. \ In this regard, their approach is more
in line with Sommerfeld's original analysis, wherein complex variables also
play a central role.

Neither of these treatments invoke Riemannian geometry as we have done here
for hyperspheres. \ However it is possible in principle to consider the
conducting disk in $\mathbb{E}_{3}$ as a squashed oblate spheroid, and thereby
obtain the Green function for the disk by taking a limit of Green functions on
branched manifolds connected by spheroidal generalizations of the Ellis
wormhole, analogous to the $p$-norm wormholes used above. \ This more
geometrical treatment will be discussed elsewhere \cite{AlshalCurtright}.

\section{Conclusions}

We have obtained the Green function for grounded hyperspheres in $N$ spatial
dimensions by first constructing Green functions on Riemannian manifolds
(rather unfortunately, in our opinion, but commonly known as \textquotedblleft
wormholes\textquotedblright) and then by squashing these manifolds to produce
two copies of flat Euclidean space creased together along a hypersphere of
radius $R$. \ The distribution of source and image charges on the final
squashed manifold illustrates Sommerfeld's generalization of
\href{https://en.wikipedia.org/wiki/William_Thomson,_1st_Baron_Kelvin}{Thomson}%
's method.

Sommerfeld knew that his generalized method could be used to solve a large
variety of problems \cite{Sommerfeld}, writing to
\href{https://en.wikipedia.org/wiki/Felix_Klein}{Klein} in the spring of 1897
(see \cite{Eckert} page 80):

\begin{quote}
\textquotedblleft The number of boundary value problems solvable by means of
my elaborated Thomson's method of images is very great.\textquotedblright
\end{quote}

\noindent But he does not seem to have pursued this during the next
half-century, perhaps because more interesting mathematics and physics
questions captured his attention.

In our opinion, the most prescient aspect of Sommerfeld's nineteenth century
work lies in its suggestion that physical problems in electromagnetic theory
may be simplified and perhaps more easily understood through the study of
Riemannian geometries, a view that developed much later in general relativity.
\ However, like \href{https://en.wikipedia.org/wiki/Bernhard_Riemann}{Riemann}
before him, in 1897 Sommerfeld had no reason to include time along with the
spatial dimensions of his envisioned manifolds, thus making his work
premature. \ 

Nevertheless, considering its application of Riemann's ideas from geometry and
complex analysis to higher dimensional branched manifolds, we believe
Sommerfeld's work should be recognized as a legitimate precursor to the
wormhole studies that appeared a few decades later
\cite{Flamm,EinsteinRosen,Ellis} and continue to the present day
\cite{MorrisThorne,JamesEtAl,Lobo}. \ We hope our paper encourages readers to
share this opinion.\bigskip

\noindent\textbf{Acknowledgements} \ It has been our pleasure to reconsider
this elementary subject during the year of the Feynman Centennial and the
Sommerfeld Sesquicentennial. \ This work was supported in part by a University
of Miami Cooper Fellowship, and by a Clark Way Harrison Visiting Professorship
at Washington University in Saint Louis.

\vfill

\section*{Figures}

The first nine Figures show equatorial surface slices of various $p$-norm
wormholes, as embeddings in three dimensions, where
\begin{equation}
\left(  ds\right)  ^{2}=\left(  dw\right)  ^{2}+r^{2}\left(  w\right)  \left(
d\theta\right)  ^{2}=\left(  dx\right)  ^{2}+\left(  dy\right)  ^{2}+\left(
dz\right)  ^{2} \tag{F1}%
\end{equation}%
\begin{equation}
x\left(  w,\theta\right)  =r\left(  w\right)  \cos\theta\ ,\ \ \ y\left(
w,\theta\right)  =r\left(  w\right)  \sin\theta\ ,\ \ \ r\left(  w\right)
=\left(  R^{p}+\left(  w^{2}\right)  ^{p/2}\right)  ^{1/p} \tag{F2}%
\end{equation}%
\begin{equation}
z\left(  w\right)  =\int_{0}^{w}\sqrt{1-\left(  dr\left(  \varpi\right)
/d\varpi\right)  ^{2}}d\varpi=\int_{0}^{w}\sqrt{1-\left(  \varpi^{2}\right)
^{p-1}\left(  R^{p}+\left(  \varpi^{2}\right)  ^{p/2}\right)  ^{\frac{2}{p}%
-2}}\,d\varpi\tag{F3}%
\end{equation}
For example, for $p=2$,%
\begin{equation}
z\left(  w\right)  =R\ln\left(  \frac{w+\sqrt{R^{2}+w^{2}}}{R}\right)
=R\operatorname{arcsinh}\left(  \frac{w}{R}\right)  \tag{F4}%
\end{equation}
For generic $p$, it is easiest to obtain $z\left(  w\right)  $ by numerical
solution of
\begin{equation}
\frac{dz\left(  w\right)  }{dw}=\sqrt{1-\left(  w^{2}\right)  ^{p-1}\left(
R^{p}+\left(  w^{2}\right)  ^{p/2}\right)  ^{\frac{2}{p}-2}} \tag{F5}%
\end{equation}
with initial condition $z\left(  0\right)  =0$. \ 

\vfill\newpage

\noindent Figures 1-9: \ Embedded $p$-norm wormhole equatorial surfaces for
$p$ as shown.\ \ All plots are for $R=1$, with $0\leq\theta\leq2\pi$ and
$-2\leq w\leq2$. \ Upper and lower branches of the surfaces are in orange and
green, respectively.

\noindent\hspace{-0.5in}%
{\parbox[b]{2.5391in}{\begin{center}
\includegraphics[
trim=0.000000in 0.000000in 0.000000in -0.685440in,
height=2.6091in,
width=2.5391in
]%
{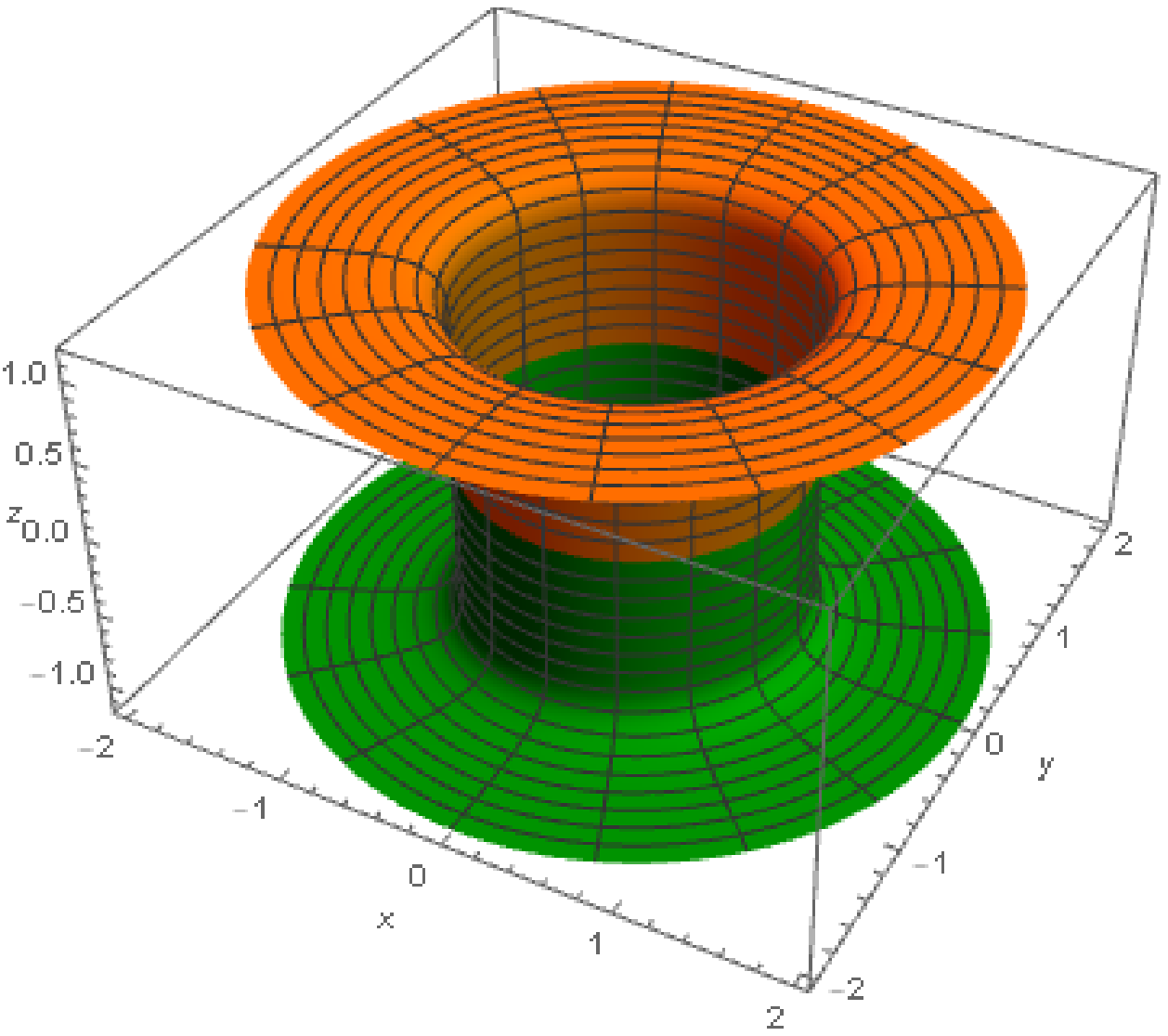}%
\\
Figure 1: \ $p=17$%
\end{center}}}%
{\parbox[b]{2.5391in}{\begin{center}
\includegraphics[
trim=0.000000in 0.000000in 0.000000in -0.490956in,
height=2.6091in,
width=2.5391in
]%
{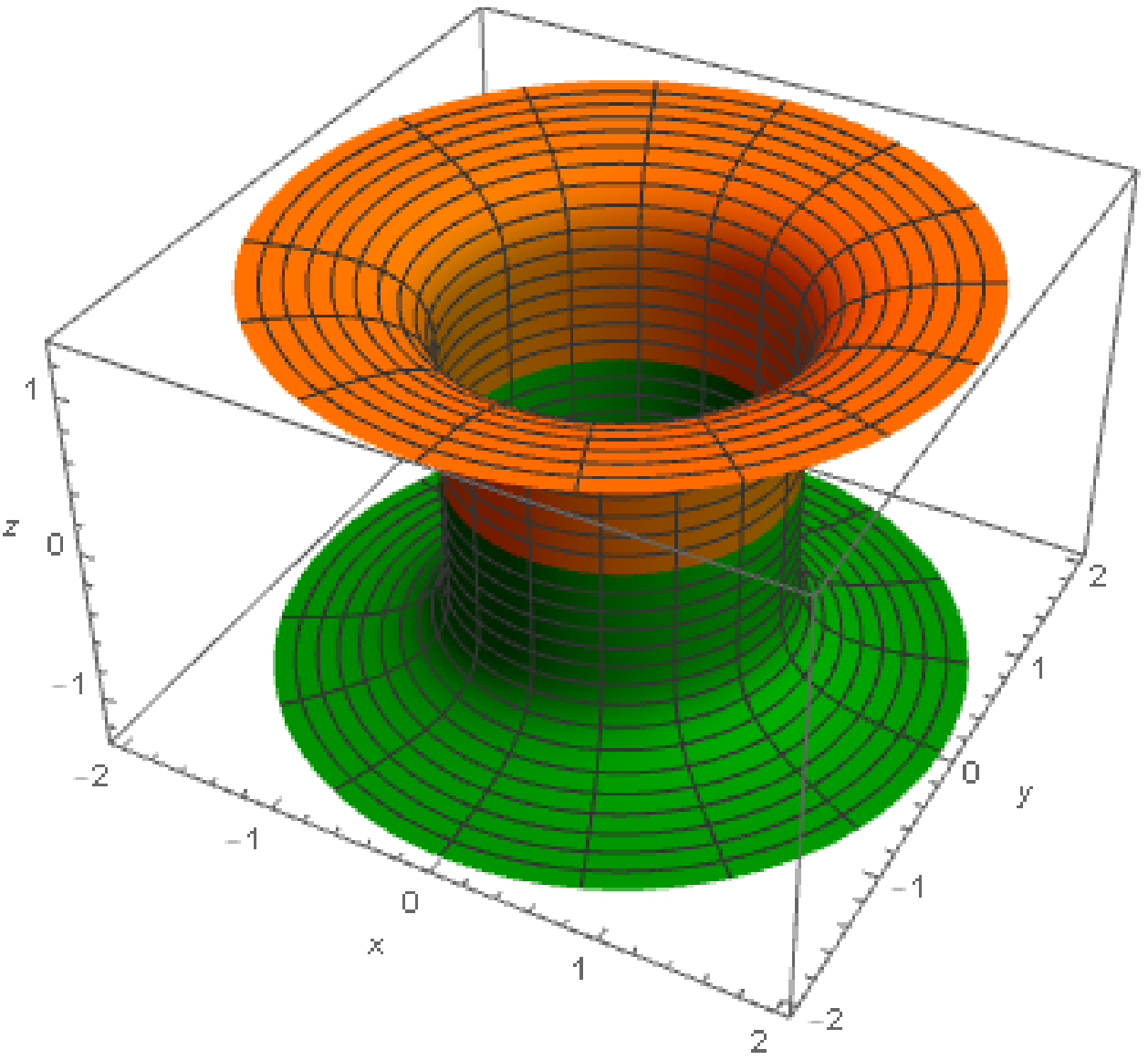}%
\\
Figure 2: $\ p=9$%
\end{center}}}%
{\parbox[b]{2.5391in}{\begin{center}
\includegraphics[
trim=0.000000in 0.000000in 0.000000in -0.110616in,
height=2.6091in,
width=2.5391in
]%
{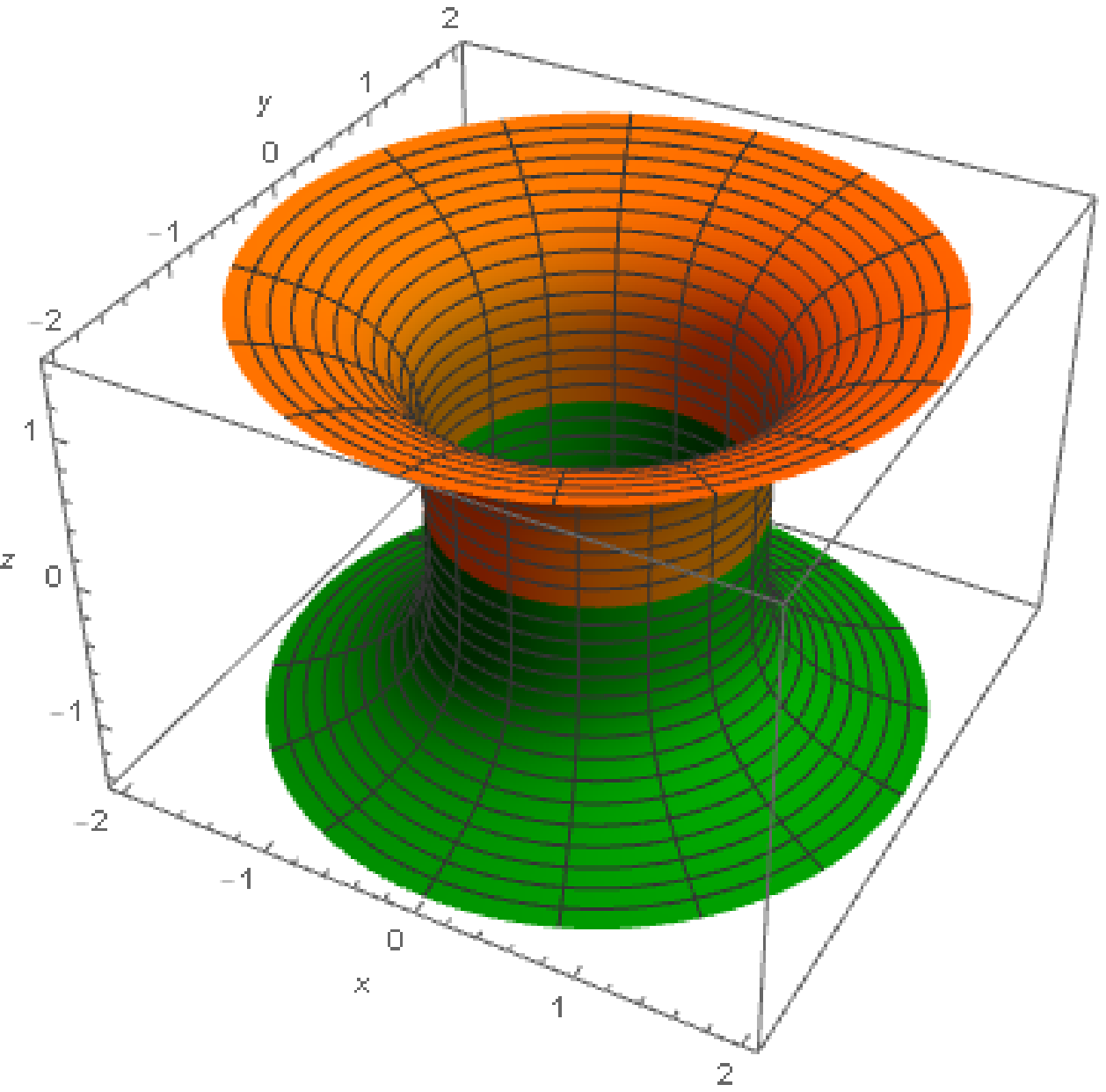}%
\\
Figure 3: $\ p=5$%
\end{center}}}%
\vspace{-0.25in}

\noindent\hspace{-0.5in}%
{\parbox[b]{2.5391in}{\begin{center}
\includegraphics[
trim=0.000000in 0.000000in 0.000000in -0.140011in,
height=2.6091in,
width=2.5391in
]%
{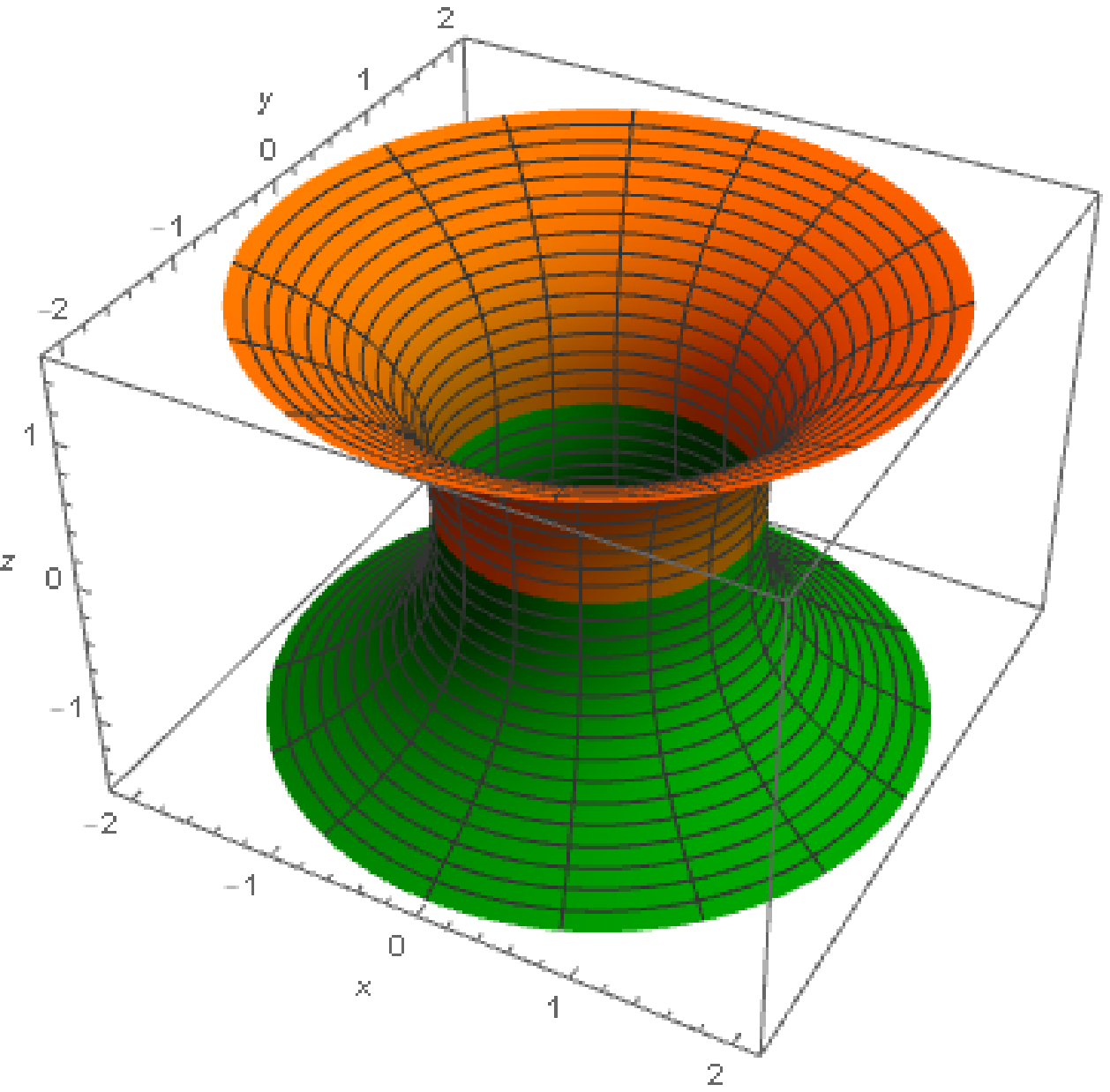}%
\\
Figure 4: $\ p=3$%
\end{center}}}%
{\parbox[b]{2.5391in}{\begin{center}
\includegraphics[
trim=0.000000in 0.000000in 0.000000in -0.340003in,
height=2.6091in,
width=2.5391in
]%
{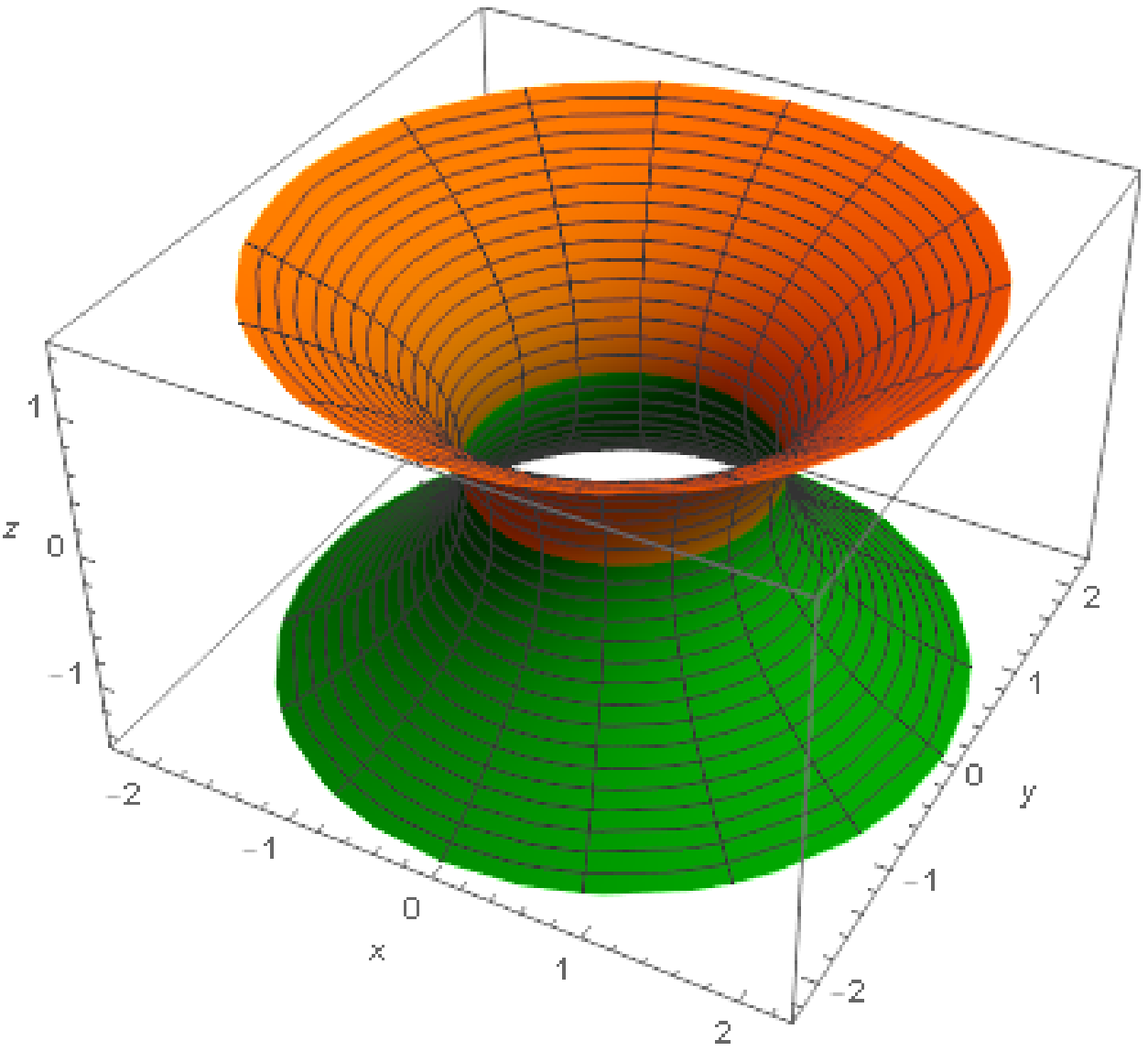}%
\\
Figure 5: $\ p=2$%
\end{center}}}%
{\parbox[b]{2.5391in}{\begin{center}
\includegraphics[
trim=0.000000in 0.000000in 0.000000in -0.685440in,
height=2.6091in,
width=2.5391in
]%
{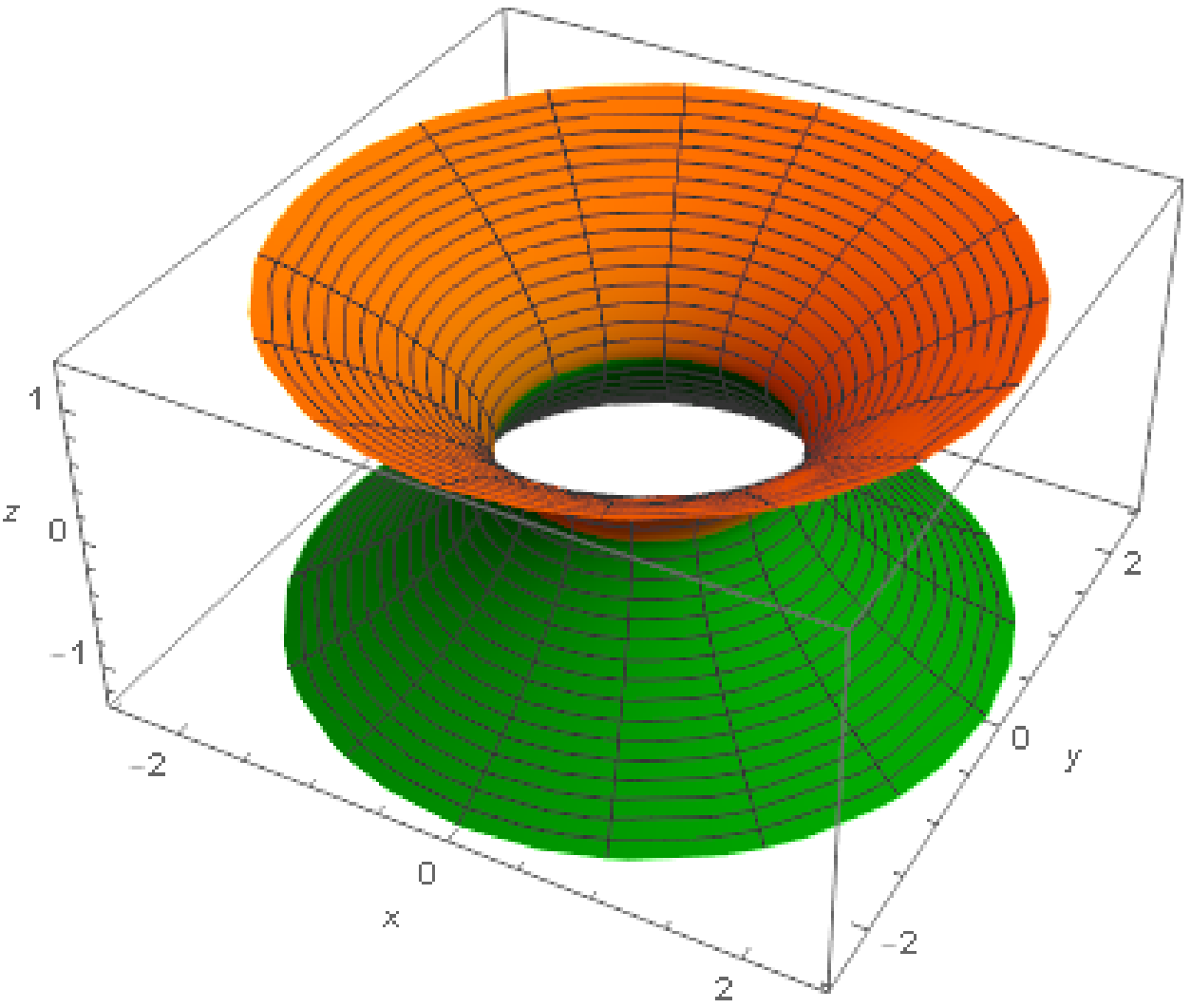}%
\\
Figure 6: \ $p=3/2$%
\end{center}}}%
\vspace{-0.25in}

\noindent\hspace{-0.5in}%
{\parbox[b]{2.5391in}{\begin{center}
\includegraphics[
trim=0.000000in 0.000000in 0.000000in -0.212188in,
height=1.9164in,
width=2.5391in
]%
{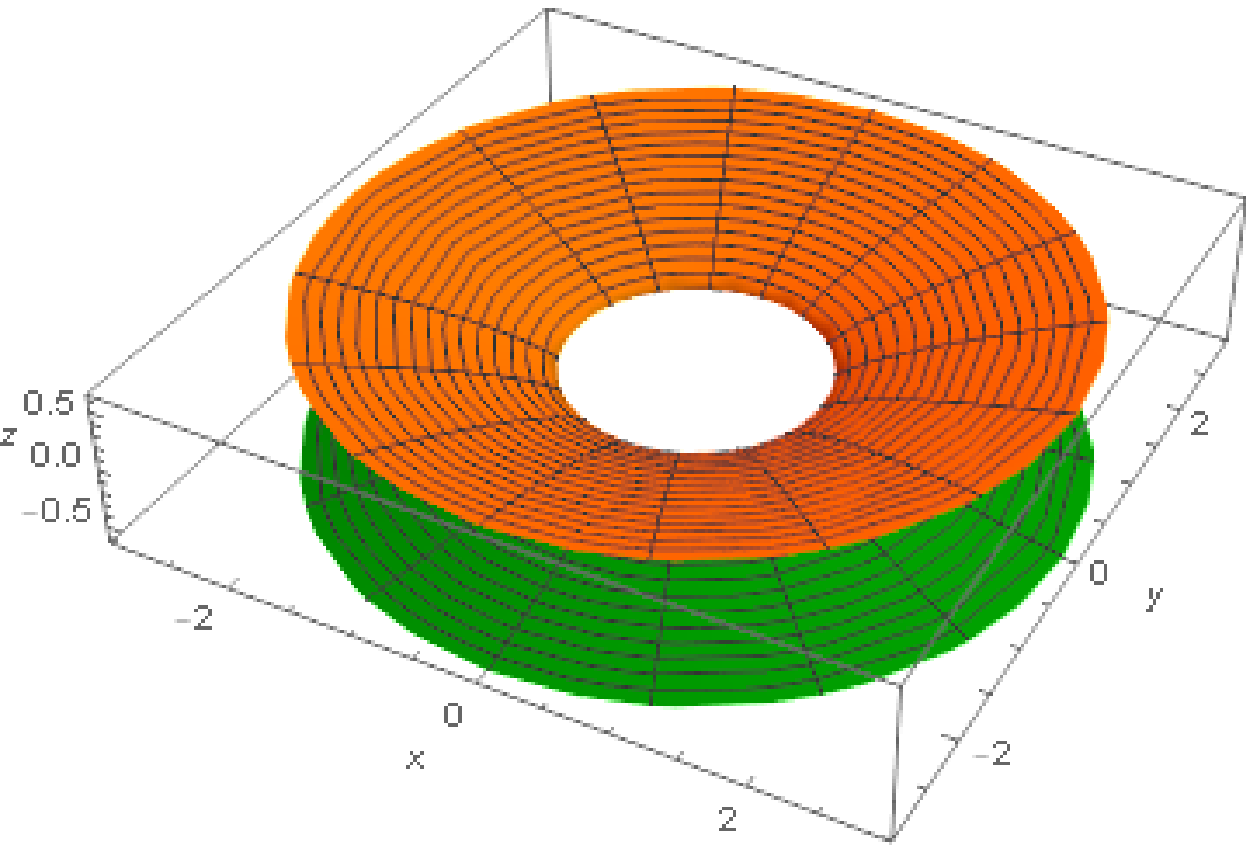}%
\\
Figure 7: $\ p=17/16$%
\end{center}}}%
{\parbox[b]{2.5391in}{\begin{center}
\includegraphics[
trim=0.000000in 0.000000in 0.000000in -0.138534in,
height=1.7469in,
width=2.5391in
]%
{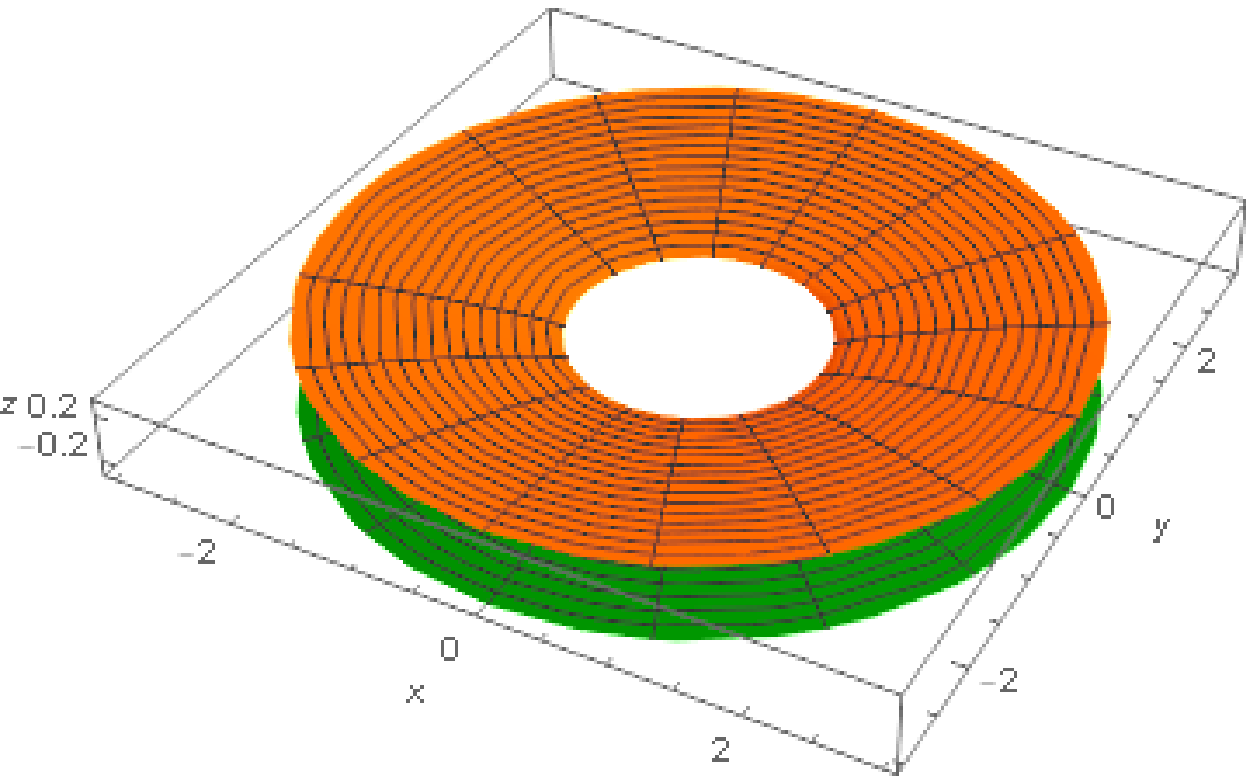}%
\\
Figure 8: $\ p=65/64$%
\end{center}}}%
{\parbox[b]{2.5391in}{\begin{center}
\includegraphics[
trim=0.000000in 0.000000in 0.000000in -0.163674in,
height=1.676in,
width=2.5391in
]%
{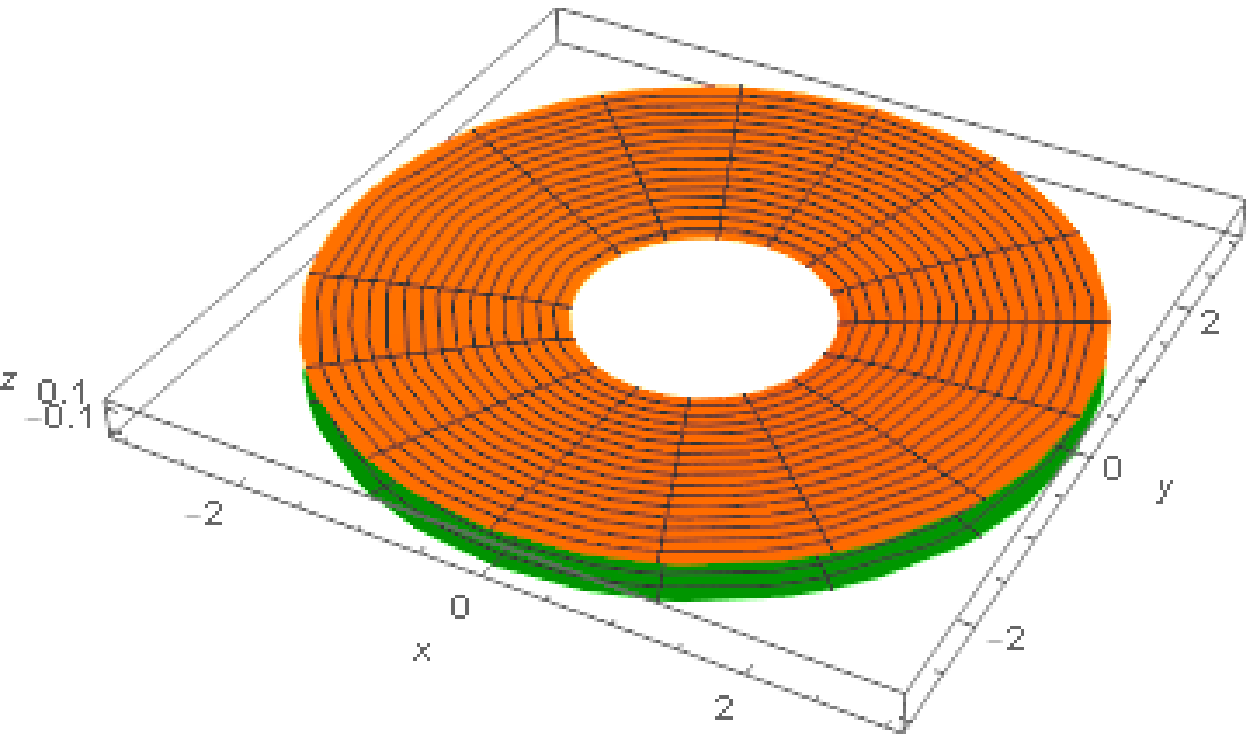}%
\\
Figure 9: $\ p=257/256$%
\end{center}}}%
\vspace{-0.25in}\newpage

\noindent The next four Figures show contour plots of various Green functions
with $N=4$ and $R=1$, for $-\pi\leq\theta\leq\pi$ along the vertical axes,
where $\theta\equiv\arccos\left(  \widehat{r_{1}}\cdot\widehat{r_{2}}\right)
$, and for $-2.5\leq w_{1}\leq2.5$ along the horizontal axes. \ The contours
show $\left\vert G\text{ or }G_{o}\right\vert \leq0.25$.\bigskip

\noindent Figure 10: \ Plot of (\ref{N=4EllisGreen}) versus $w_{1}$ and
$\theta$ with source at $\left(  w_{2},\widehat{r_{2}}\right)  =\left(
1,\widehat{r_{2}}\right)  $.\bigskip

\noindent Figure 11: \ Plot of (\ref{GoGreenFcn}) using (\ref{N=4EllisGreen}%
)\ with source at $\left(  w_{2},\widehat{r_{2}}\right)  =\left(
1,\widehat{r_{2}}\right)  $ \& image at $\left(  -w_{2},\widehat{r_{2}%
}\right)  =\left(  -1,\widehat{r_{2}}\right)  $.\bigskip

\noindent Figure 12: \ Plot of (\ref{Gflat++}) and (\ref{Gflat+-}) for $N=4$
versus $w_{1}$ and $\theta$ with source at $\left(  w_{2},\widehat{r_{2}%
}\right)  =\left(  1,\widehat{r_{2}}\right)  $.\bigskip

\noindent Figure 13: \ Plot of (\ref{GreenGroundedSquashedUpper}) and
(\ref{GreenGroundedSquashedLower})\ for $N=4$, with source at $\left(
w_{2},\widehat{r_{2}}\right)  =\left(  1,\widehat{r_{2}}\right)  $ \& image at
$\left(  -w_{2},\widehat{r_{2}}\right)  =\left(  -1,\widehat{r_{2}}\right)
$.\bigskip

\begin{center}%
{\parbox[b]{2.5391in}{\begin{center}
\includegraphics[
trim=0.000000in 0.000000in 0.000000in -0.765532in,
height=2.7899in,
width=2.5391in
]%
{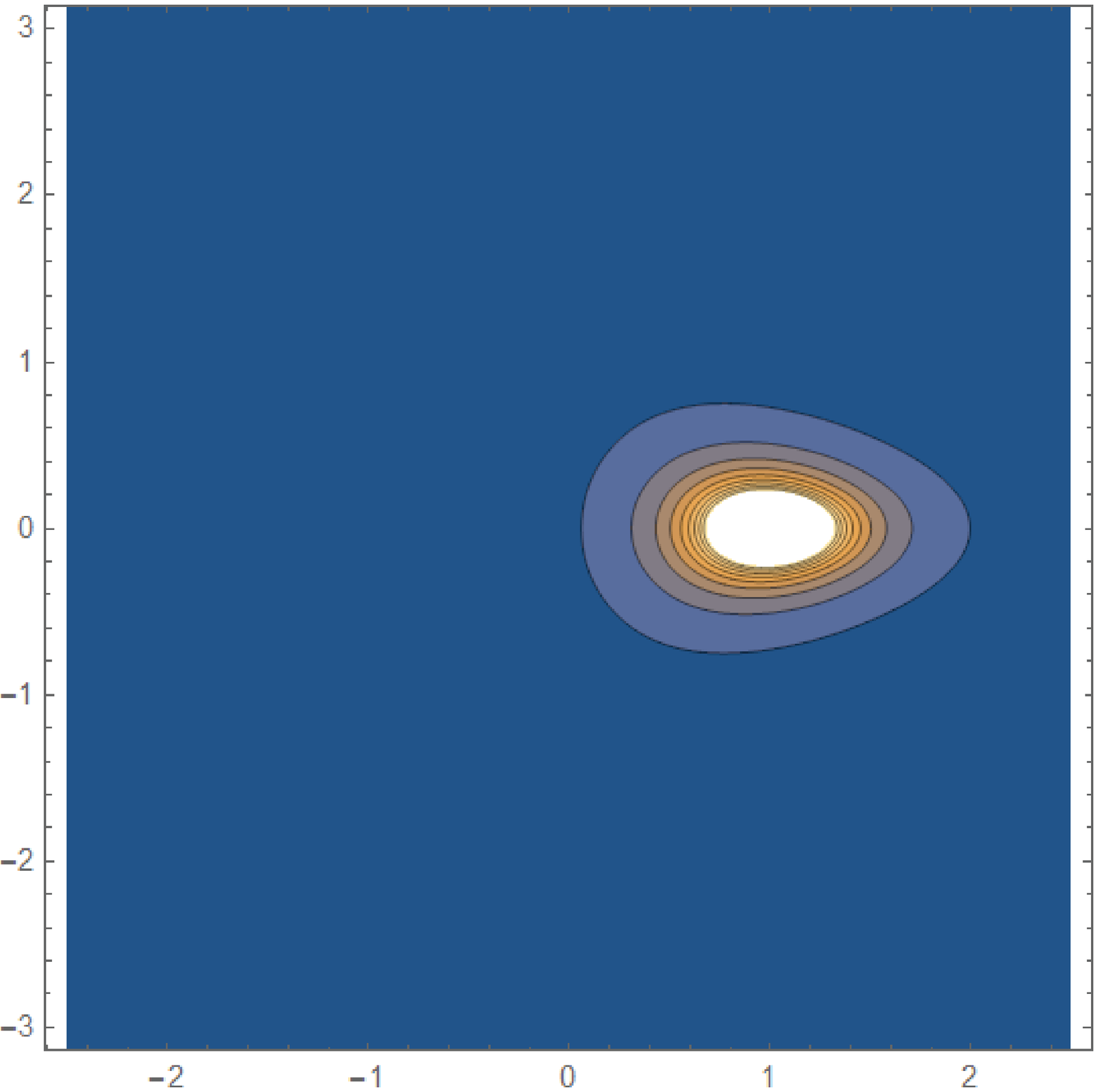}%
\\
Figure 10: \ Contour plot of $G$ for the Ellis wormhole in 4D.
\end{center}}}%
\ \ \ \
{\parbox[b]{2.5391in}{\begin{center}
\includegraphics[
trim=0.000000in 0.000000in 0.000000in -0.765532in,
height=2.7899in,
width=2.5391in
]%
{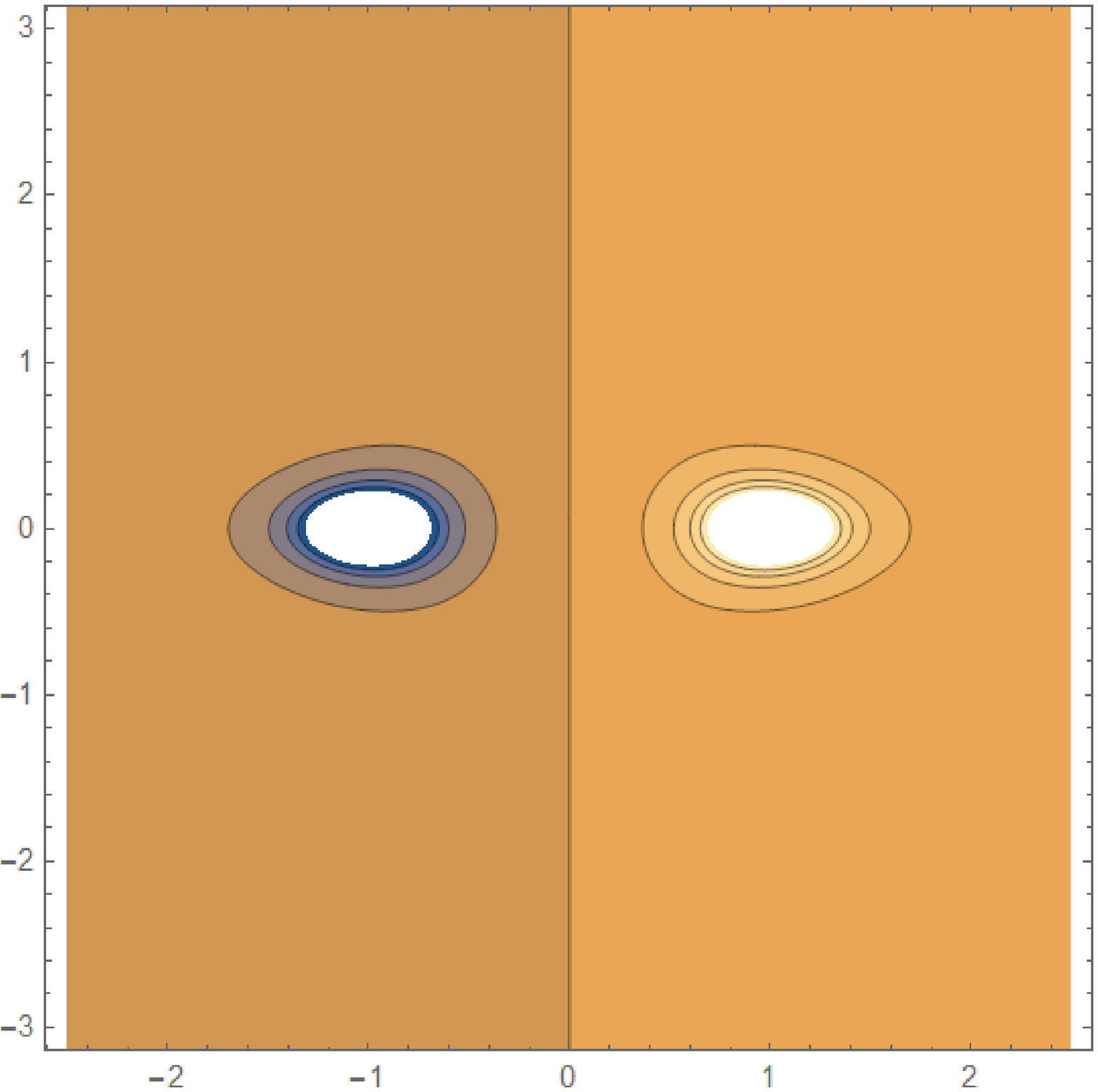}%
\\
Figure 11: $\ $Contour plot of $G_{o}$ for the Ellis wormhole in 4D.
\end{center}}}%

\bigskip%

{\parbox[b]{2.5391in}{\begin{center}
\includegraphics[
trim=0.000000in 0.000000in 0.000000in -0.765532in,
height=2.7899in,
width=2.5391in
]%
{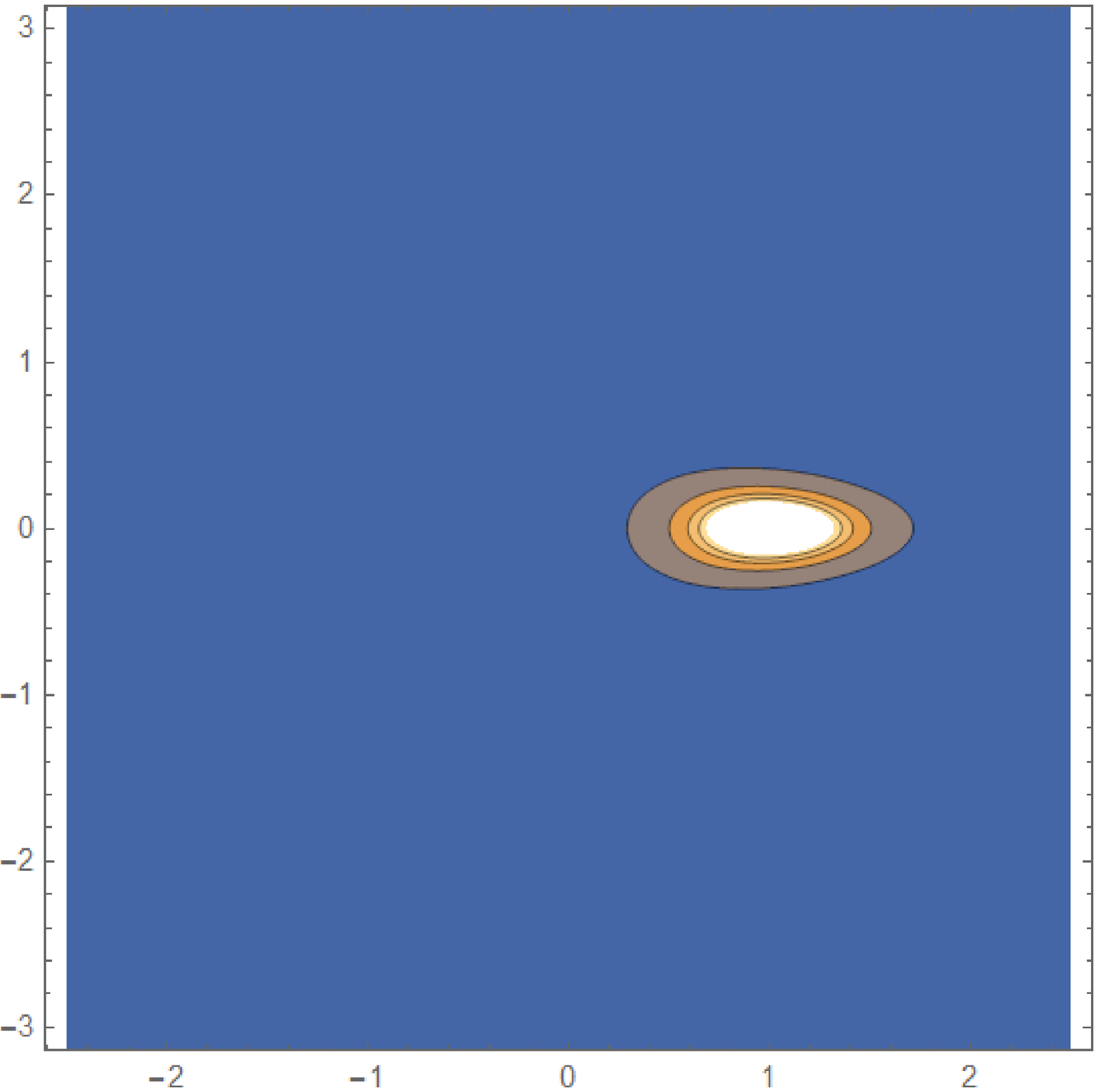}%
\\
Figure 12: $\ $Contour plot of $G$ for the squashed wormhole in 4D.
\end{center}}}%
\ \ \ \ \
{\parbox[b]{2.5391in}{\begin{center}
\includegraphics[
trim=0.000000in 0.000000in 0.000000in -0.765532in,
height=2.7899in,
width=2.5391in
]%
{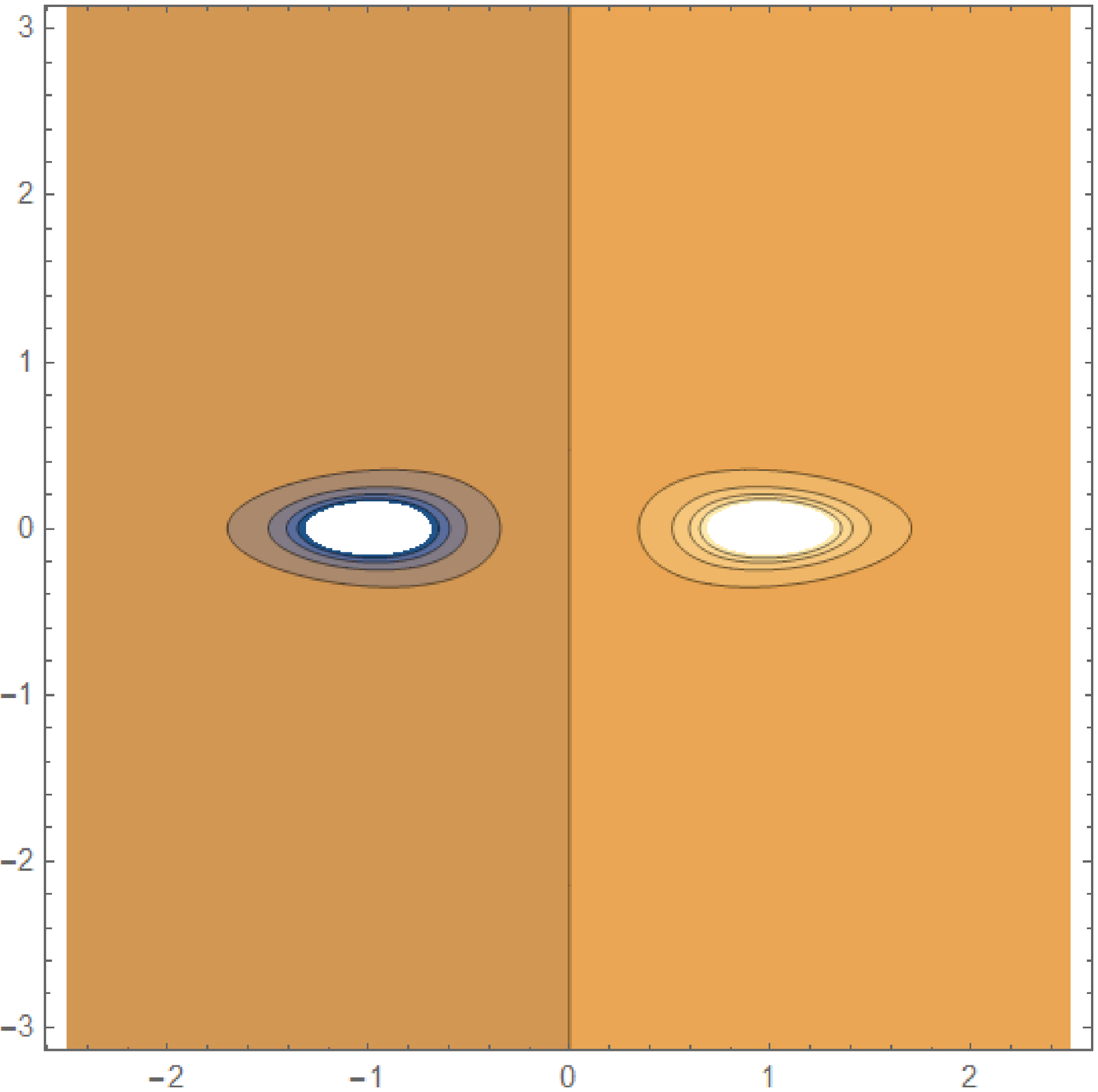}%
\\
Figure 13: $\ $Contour plot of $G_{o}$ for the squashed wormhole in 4D.
\end{center}}}%

\end{center}


\bigskip

\begin{thebibliography}{99}                                                                                               %


\bibitem {AlshalCurtright}H Alshal and T Curtright, in preparation.

\bibitem {CurtrightEtAl}T Curtright, H Alshal, P Baral, S Huang, J Liu, K
Tamang, X Zhang, and Y Zhang, \textquotedblleft The Conducting Ring Viewed as
a Wormhole\textquotedblright%
\ [\href{https://arxiv.org/abs/1805.11147}{https://arxiv.org/abs/1805.11147}].

\bibitem {DavisReitz}L C Davis and J R Reitz, \textquotedblleft Solution to
potential problems near a conducting semi-infinite sheet or conducting
disc\textquotedblright\ \href{https://doi.org/10.1119/1.1976616}{Am. J. Phys.
39 (1971) 1255-1265}.

\bibitem {DavisReitzAgain}L C Davis and J R Reitz, \textquotedblleft Solution
of potential problems near the corner of a conductor\textquotedblright%
\ \href{https://doi.org/10.1063/1.522671}{J. Math. Phys. 16 (1975) 1219--1226}.

\bibitem {Duffy}D G Duffy,
\textit{\href{https://www.crcpress.com/Greens-Functions-with-Applications-Second-Edition/Duffy/p/book/9781138894464}{\textit{Green's
Functions with Applications}}}, Second Edition, CRC Press (2017) ISBN-13: 978-1482251029.

\bibitem {Eckert}M Eckert,
\textit{\href{https://www.amazon.com/Arnold-Sommerfeld-Science-Turbulent-1868-1951/dp/1461474604/ref=sr_1_7?s=books&ie=UTF8&qid=1524241334&sr=1-7&keywords=sommerfeld}{\textit{Arnold
Sommerfeld: \ Science, Life and Turbulent Times 1868-1951}}}, Springer-Verlag
(2013) ISBN-13: 978-1461474609.

\bibitem {EinsteinRosen}A Einstein and N Rosen, \textquotedblleft The Particle
Problem in the General Theory of Relativity\textquotedblright%
\ \href{https://dx.doi.org/10.1103/PhysRev.48.73}{Phys. Rev. 48 (1935) 73-77}.

\bibitem {Ellis}H G Ellis, \textquotedblleft Ether flow through a drainhole:
\ A particle model in general relativity\textquotedblright%
\ \href{https://doi.org/10.1063/1.1666161}{J. Math. Phys. 14 (1973) 104--118}.

\bibitem {Flamm}L Flamm, \textquotedblleft Beitr\"{a}ge zur Einsteinschen
Gravitationstheorie\textquotedblright%
\ \href{https://babel.hathitrust.org/cgi/pt?id=mdp.39015010783705;view=1up;seq=484}{Physikalische
Zeitschrift 17 (1916) 448-454}.

\bibitem {Hobson}E. W. Hobson, \textquotedblleft On Green's function for a
circular disc, with application to electrostatic problems\textquotedblright%
\ \href{https://babel.hathitrust.org/cgi/pt?id=coo.31924069328965;view=1up;seq=315}{Trans.
Cambridge Philos. Soc. 18 (1900) 277- 291}.

\bibitem {JamesEtAl}O James, E von Tunzelmann, P Franklin, and K S Thorne,
\textquotedblleft Visualizing \textit{Interstellar}'s
Wormhole\textquotedblright\ \href{https://doi.org/10.1119/1.4916949}{Am. J.
Phys. 83 (2015) 486-499}
[\href{https://arxiv.org/abs/1502.03809}{https://arxiv.org/abs/1502.03809}].

\bibitem {Lobo}F S N Lobo (editor),
\textit{\href{https://www.amazon.com/Wormholes-Conditions-Fundamental-Theories-Physics/dp/3319551817/ref=sr_1_14?s=books&ie=UTF8&qid=1527624030&sr=1-14&keywords=wormholes}{\textit{Wormholes,
Warp Drives and Energy Conditions}}}, Springer-Verlag (2017) ISBN-13: 978-3319551814.

\bibitem {Mehra}J Mehra,
\textit{\href{http://www.amazon.com/Beat-Different-Drum-Science-Richard/dp/0198539487/ref=sr_1_3?s=books&ie=UTF8&qid=1456023383&sr=1-3&keywords=mehra}{\textit{The
Beat of a Different Drum: \ The Life and Science of Richard Feynman}}%
},\textit{ }Oxford University Press (1994) ISBN-13: 978-0198539483.

\bibitem {MorrisThorne}M S Morris and K S Thorne, \textquotedblleft Wormholes
in spacetime and their use for interstellar travel: \ A tool for teaching
general relativity\textquotedblright%
\ \href{https://dx.doi.org/10.1119/1.15620}{Am. J. Phys. 56 (1988) 395-412}.

\bibitem {Sommerfeld}A Sommerfeld, \textquotedblleft\"{U}ber verzweigte
Potentiale im Raum\textquotedblright%
\ \href{https://doi.org/10.1112/plms/s1-28.1.395}{Proc. London Math. Soc.
(1896) s1-28 (1): 395-429}; ibid. 30 (1899) 161.

\bibitem {Waldmann}L Waldmann, \textquotedblleft Zwei Anwendungen der
Sommerfeld'schen Methode der verzweigten Potentiale\textquotedblright%
\ \href{https://babel.hathitrust.org/cgi/pt/search?id=mdp.39015076063125;view=1up;seq=5;q1=Waldmann;start=1;sz=10;page=search;orient=0}{Physikalische
Zeitscrift 38 (1937) 654--663}.
\end{thebibliography}
\end{document}